\documentclass[12pt,preprint]{aastex}

\newcommand{\kmpers}{{{\rm \; km\;s}^{-1}}}

\newcommand{\dex}{{\rm \; dex}}
\newcommand{\magnitude}{{\rm \; mag}}

\newcommand{\logFeH}{{\log{{\rm(Fe/H)}}}}
\newcommand{\logFeIIH}{{\log{{\rm(Fe \; II/H)}}}}

\newcommand{\Htot}{{\rm H_{tot}}}

\newcommand{\NHI}{{N{\rm(H \; I)}}}

\newcommand{\NHmol}{{N\rm(H_{2})}}
\newcommand{\NFe}{{N{\rm(Fe)}}}
\newcommand{\NFeII}{{N{\rm(Fe \; II)}}}
\newcommand{\logFeII}{{\log{N{\rm(Fe \; II)}}}}

\newcommand{\logHI}{{\log{\NHI}}}
\newcommand{\logHmol}{{\log{\NHmol}}}
\newcommand{\logHtot}{{\log{\NHtot}}}

\newcommand{\NHtot}{{N{\rm(H_{tot})}}}
\newcommand{\Hmol}{{\rm H_{2}}}

\newcommand{\nHavg}{{< \! \! \rm{n_{H}} \! \! >}}

\newcommand{\lognHavg}{{\log{\nHavg}}}

\newcommand{\fHmol}{{f\rm(H_{2})}}
\newcommand{\eqw}{{W_{\lambda}}}
\newcommand{\av}{{A_V}}

\newcommand{\ebv}{{E_{B-V}}}

\newcommand{\avdist}{{\av / r}}
\newcommand{\ebvdist}{{\ebv / r}}
\newcommand{\rv}{{R_V}}

\shorttitle{}

\shortauthors{Jensen and Snow}

\begin{document}

\title{NEW INSIGHTS ON INTERSTELLAR GAS-PHASE IRON}

\affil{Center for Astrophysics and Space Astronomy}
\affil{University of Colorado at Boulder, Campus Box 389}
\affil{Boulder, CO 80309-0389}

\email{Adam.Jensen@colorado.edu, tsnow@casa.colorado.edu}

\author{Adam G. Jensen and Theodore P. Snow}

\begin{abstract}
In this paper, we report on the gas-phase abundance of singly-ionized iron (Fe II) for 51 lines of sight, using data from the {\it Far Ultraviolet Spectroscopic Explorer} ({\it FUSE}).  Fe II column densities are derived by measuring the equivalent widths of several ultraviolet absorption lines and subsequently fitting those to a curve of growth.  Our derivation of Fe II column densities and abundances creates the largest sample of iron abundances in moderately- to highly-reddened lines of sight explored with {\it FUSE}, lines of sight that are on average more reddened than lines of sight in previous {\it Copernicus} studies.  We present three major results.  First, we observe the well-established correlation between iron depletion and $\nHavg$ and also find trends between iron depletion and other line of sight parameters (e.g.~$\fHmol$, $\ebv$, and $\av$), and examine the significance of these trends.  Of note, a few of our lines of sight probe larger densities than previously explored and we do not see significantly enhanced depletion effects.  Second, we present two detections of an extremely weak Fe II line at 1901.773 \AA{} in the archival STIS spectra of two lines of sight (HD 24534 and HD 93222).  We compare these detections to the column densities derived through {\it FUSE} spectra and comment on the line's $f$-value and utility for future studies of Fe II.  Lastly, we present strong anecdotal evidence that the Fe II $f$-values derived empirically in through {\it FUSE} data are more accurate than previous values that have been theoretically calculated, with the probable exception of $f_{1112}$.
\end{abstract}


\section{INTRODUCTION AND BACKGROUND}
\label{s:FeII_intro}
Iron is both a relatively abundant element in the Galaxy and a major constituent in most grain models.  Fe I has an ionization potential of only 7.87 eV, while Fe II has an ionization potential of 16.18 eV.  Therefore, the dominant form of gas-phase interstellar iron found in H I regions should be Fe II.  In H II regions, iron should be found as a mix of Fe II, Fe III, and Fe IV.  In $\Hmol$ regions, some Fe I may exist.  The average gas-phase abundance of iron that has been established implies that the vast majority of interstellar iron is tied up in dust.  However, observed variations in the iron abundance, while having only marginal implications for the dominant ingredients of dust, may still shed light on the physical conditions of interstellar clouds.

Three major studies of interstellar gas-phase iron have been carried out: \citet{SavageBohlin} and \citet*{Jenkins1986} with {\it Copernicus} data; and \citet*{SRF2002} with {\it FUSE} data.  (For the rest of this paper, these three papers will be referred to as SB1979, JSS1986, and SRF2002, respectively.)  \citet{Howk} also studied Fe II with {\it FUSE} but for the purpose of empirically determining oscillator strengths ($f$-values) in the {\it FUSE} wavelength region rather than studying abundances and depletions.

In summarizing previous studies and discussing our own results, it will be useful to define a few terms.  By abundance, we mean the ratio of an element relative to hydrogen, given either logarithmically or linearly.  We will define the iron abundance as Fe II/$\Htot=\NFeII/[\NHI + 2\NHmol]$, and neglect other ionization states of iron and hydrogen for the reasons discussed above.  The quantity $\delta$ is the ratio of the gas-phase abundance seen in a particular line of sight to an assumed cosmic abundance standard.  Despite difference in terminology in the literature (cf.~SB1979 and SRF2002), we will refer to the quantity $D=-\log{\delta}$ as the depletion.  It is the logarithm of the ratio of the total amount of an element to the amount in the gas-phase.  Therefore, an increase in depletion means an increase in the amount of that element tied up in dust grains and/or molecules.  This definition of depletion is useful because errors in the assumed cosmic standard produce a constant shift in the quantity $D$ and trends can still be analyzed.

SB1979 found the interstellar ratio of iron to hydrogen to be equal to $4.5\times10^{-7}$.  They observed depletions in the range of $D=0.11-2.49$.  SB1979 found positive correlations between $\NFeII$ and the quantities $\NHtot$ and $\ebv$.  These results do not consider depletion, and could be consistent with constant depletion.  However, they did find iron depletion to be correlated, to varying degrees, with several parameters.  The first very rough correlation they cite is between depletion and $E_{1330-V}/\ebv$, a measure of dust grain size.  SB1979 suggested that this implies that small grains are removed through grain growth and coalescence as opposed to destruction.  However, other dust-related parameters such as visual polarization from grains and the gas-to-dust ratio (as measured by $\NHtot / \ebv$) were not found to correlate with iron depletion.  Clearer correlations did exist for $\nHavg$ ($\equiv \NHtot/r$, where $r$ is the line-of-sight pathlength) and $\ebv/r$, two quantities that should show similar correlations because of the correlation between $\NHtot$ and $\ebv$ \citep{Bohlin1978}.  In these correlations, iron depletion increases with increased average density in the line of sight.  Two other measures of cloud conditions that SB1979 found to be correlated to iron depletion were the molecular fraction of hydrogen, $\fHmol$---positively correlated with depletion---and kinetic temperatue, $T_K$---negatively correlated with depletion.  These correlations were somewhat rough, however.  All of these correlations ($\nHavg$, $\ebv/r$, $\fHmol$, and $T_K$) imply that iron depletion increases in denser environments, possibly due to density-dependent formation or destruction processes or shielding from $\Hmol$.  Finally, SB1979 also found a correlation between the depletions of Fe and Ti, a result that, when combined with other results from the literature \citep{deBoerLamers, Stokes}, suggests similarities between grains that contain Fe, Ca, Mn, and Ti.

JSS1986 measured the abundances of several refractory elements (Mg II, P II, Cl I, Cl II, Mn II, Fe II, Cu II, and Ni II) with {\it Copernicus}.  The major conclusion of the JSS1986 study was that depletions of all the elements are strongly correlated with the average hydrogen density in the line of sight, $\nHavg$, with increasing depletion related to increased density.  Depletions of various elements are also correlated with each other.  JSS1986 interpreted these results largely in the context of the model of \citet{Spitzer1985}, where the ISM is considered to be comprised of two types of clouds, warm and cold.  According to this model, individual lines of sight should exhibit properties that are a result of sampling both types of clouds, and the best parameter for characterizing whether warm or cold clouds are sampled is $\nHavg$.  JSS1986 found $\logFeIIH=-6.46\pm0.06$ for cold clouds and $\logFeIIH=-5.84\pm0.05$ for warm clouds.  The observed random deviations from a smooth trend in the relationship with $\nHavg$ were larger than those expected from just the errors in measuring the iron and hydrogen column densities; JSS1986 used this fact to estimate that the true dispersion in the iron abundance is between 0.03 and 0.11 dex.  It is worth noting that JSS1986 expressed two observational concerns that may have affected their data:  (1) the possibility of ``hidden,'' highly-saturated components that, though undetected, might contribute significantly to the measured column densities and (2) significant contributions to the column densities from H II regions that violate the assumption that the true iron abundance, $\NFe/\NHtot$, is accurately represented by $\NFeII / [\NHI + \NHmol]$.

SRF2002 examined Fe II abundances with {\it FUSE}, and explored 18 reddened lines of sight with greater average extinction, reddening, and molecular fraction of hydrogen than the SB1979 and JSS1986 surveys, though the SRF2002 sample did not probe lines of sight with values of $\nHavg$ that were significantly larger than the densest lines of sight in SB1979 and JSS1986.  SRF2002 found that while Fe II depletion increases with increased $\ebv$ up to about $\ebv \approx 0.35\magnitude$, primarily in the {\it Copernicus} targets, the trend levels off at larger values of $\ebv$ (see Fig. 3 of SRF2002).  No correlation was found between depletion and $\av$ (Fig. 4 of SRF2002), for a wide range of extinction ($\av \approx 1-3.5\magnitude$); in fact, the depletions appear to be very constant.  SRF2002 similarly concluded that there is little correlation between iron depletion and the molecular fraction of hydrogen (Fig. 5 of SRF2002, including the results of SB1979).  Lastly, Fig. 6 of SRF2002 shows the correlation of iron depletion with average line of sight density, $\nHavg$.  That figure shows that the SRF2002 results are consistent with the correlation already observed by SB1979 and JSS1986, though, as stated above, SRF2002 did not probe larger $\nHavg$ than those surveys.  SRF2002 discussed in detail potential interpretations of the observed correlations (or lack thereof) and the fact that significantly enhanced depletions are $not$ observed in clouds with large $\ebv$, $\av$, and $\fHmol$, contrary to what was expected prior to the survey.  That discussion involved the creation and destruction of grain mantles, the warm and cold cloud mixing model of the ISM suggested by \citet{Spitzer1985}, and the definition of {\it translucent lines of sight} as opposed to {\it translucent clouds}.  This latter idea is also discussed in \citet{SnowMcCall}, with the principle being that diffuse material can make a significant contribution to the reddening and extinction in a line of sight that might otherwise be assumed to match the assumed characteristics of a ``translucent cloud.''  We will return to some of these ideas in \S \ref{s:FeII_results}.

A more recent but smaller survey by \citet{Miller2007} examined the iron and silicon abundances in six lines of sight characterized as ``translucent.''  \citeauthor{Miller2007}~found that both iron and silicon depletion increases in lines of sight with increased values of the extinction parameter $c_4$ \citep[in the scheme of][]{FM1988}.  Larger values of $c_4$ should correspond to an increase in the population of small grains.  Therefore, \citet{Miller2007} suggest that when additional depletion of iron and silicon is observed, those elements have been incorporated into smaller grains.  It is important to note that \citet{Miller2007} used weak lines in the STIS wavelength region to measure the Fe II abundance.  Four of the six lines of sight from \citet{Miller2007} have also been observed with {\it FUSE} and have measurements of the iron abundance found in either SRF2002 (HD 27778 and HD 207198) or this paper (HD 147888 and HD 152590).

We have carried out this study for the following reasons (roughly in decreasing order of importance):

\begin{enumerate}
\item To increase the sample size of reddened lines of sight with measurements of iron abundances and depletions.  All of our lines of sight have $\logHtot \gtrsim 21$.  Additionally, $\sim \frac{2}{3}$ of our lines of sight are characterized by $\av \gtrsim 1$,  $\sim \frac{2}{3}$ by $\ebv \gtrsim 0.3$, and  $\sim \frac{2}{3}$ by $\fHmol \gtrsim 0.1$.  In conjunction with the 16 unique lines of sight from SRF2002---we have reanalyzed HD 24534 and HD 73882 in this paper---our data nearly quadruples the size of the overall sample of reddened lines of sight examined with {\it FUSE}.  Additionally, while most of the average line of sight densities in our sample are not significantly denser than the densities in previous studies, two lines of sight, HD 147888 and HD 179406, do probe larger $\nHavg$ than any line of sight in SRF2002.
\item To probe lines of sight with less extinction and/or reddening than SRF2002.  This is useful because certain extinction parameters (specifically $\av$ and $\rv$) were not available in many cases for {\it Copernicus} targets (cf.~Figure 4 of SRF2002).  In addition, many {\it Copernicus} targets were short-pathlength lines of sight toward bright stars; using {\it FUSE}, we are able to probe fainter targets and longer pathlengths (though there is some overlap between the two samples).
\item To present two detections of a very weak line of Fe II ($\lambda_{\rm rest}$=1901.773 \AA{}) in archival STIS spectra (lines of sight toward HD 24534 and HD 93222).  We place limits on the $f$-value of this line and discuss its utility for future determinations of Fe II.
\item To evaluate the $f$-values of the Fe II used in this study, comparing the empirical values from \citet{Howk} and the theoretical calculations of \citet{RU1998}.
\item To make simple models of Fe II in several lines of sight where we are also studying the 1260 \AA{} line (along with other absorption lines) of Si II (A. G. Jensen \& T. P. Snow, in preparation).
\end{enumerate}

In \S \ref{s:FeII_obsdata} we discuss our observations and data reduction, including comments on the absorption lines used and our measurement methods.  In \S \ref{s:methods} we discuss our methods of deriving column densities, abundances, and depletions.  In \S \ref{s:FeII_results} we discuss our results and in \S \ref{s:FeII_summary} we summarize.

\section{OBSERVATIONS AND DATA REDUCTION}
\label{s:FeII_obsdata}
Archival data from both {\it FUSE} and {\it HST} is available for all 51 of the lines of sight in this study.  Lines of sight in SRF2002 were taken from the translucent cloud survey of \citet{Rachford2002}.  Our lines of sight are from a combination of the following sources:  (1) \citet{Rachford2002}---lines of sight not analyzed in SRF2002; (2) ongoing translucent cloud surveys (B. L. Rachford et al., in preparation); and (3) a Galactic disk survey of $\Hmol$ (J. M. Shull et al., in preparation).  Two of the lines of sight in this study were previously analyzed by SRF2002 (HD 24534 and HD 73882).  The HD 24534 line of sight has been reanalyzed because in it we detect the weak 1901.773 \AA{} line of Fe II.


Measurements of H I and $\Hmol$ are available (in the literature, via private communication, or by our own measurements or estimations) for all of these lines of sight.  We use the derived values of $\NFeII$ and $\NHtot$---the latter of which, as discussed in \S \ref{s:FeII_intro}, we define as $\NHI + 2\NHmol$, neglecting contributions from H II---to examine the Fe/H abundance.  We also take values from the literature of parameters such as selective extinction $\ebv$, total extinction $\av$, and distance $r$.  In a few cases, we make our own derivations of these quantities.  We then examine Fe II depletions for correlations with these or other calculated parameters (e.g.~average hydrogen volume density, $\nHavg \equiv \NHtot/r$, and the ratio of total to selective extinction, $\rv \equiv \av/\ebv$).

A summary of our program stars is found in Table \ref{stellardata}.  The atomic and molecular hydrogen content of these lines of sight is found in Table \ref{hydrogentable} and discussed in \S \ref{ss:FeII_hydrogen}.  Extinction parameters are summarized in Table \ref{reddeningtable}.  The rest of this section describes {\it FUSE} and {\it HST} data and the observation of the UV absorption lines of Fe II.

\subsection{{\it FUSE} Data}
\label{ss:FUSEdata}
Data are collected with the {\it FUSE} satellite by four channels, two channels with lithium fluoride coatings on the aluminum reflecting surfaces (LiF1 and LiF2) and two channels with reflecting surfaces of silicon carbide (SiC1 and SiC2).  Each channel contains two data segments, designated as A and B, covering adjacent wavelength regions.  The lithium fluoride channels cover the wavelength region from 989-1188 \AA{}, while the silicon carbide channels cover the region from 905-1104 \AA{}.  For more information on {\it FUSE}, see \citet{Moos2002} and \citet{Sahnow}.

The data were reduced with the CALFUSE pipeline, versions 2.4 or 3.1.  While data from different data segments are usually not coadded (e.~g.~\citeauthor{JensenOI}~\citeyear{JensenOI}, \citeyear{JensenNI}; SRF2002) we have chosen to coadd the data segments in this study to boost signal-to-noise (S/N).  By examining data segments individually we find that equivalent width measurements of absorption lines in the the coadded spectra are consistent with those from the source spectra but with smaller errors; therefore, we use the results of the coadded spectra.

The resolution element of {\it FUSE} is $\sim15-20\kmpers$.  In fitting the non-Gaussian profiles exhibited by some of the lines in this study, we used a Gaussian point-spread function (PSF) with full-width at half-maximum (FWHM) of $15\kmpers$.  Our primary objective is to determine the equivalent widths of the various absorption lines (as opposed to directly deriving column densities or velocity dispersions, i.e.~$b$-values, from individual absorption lines); therefore, the results of our fits do not depend strongly on the exact PSF that is used.

\subsection{{\it HST} Data}
\label{ss:FeII_HSTdata}
The STIS instrument onboard {\it HST} does not cover the wavelength range of the FUV absorption lines used in this study; however, STIS does cover wavelengths near 1900 \AA{}.  There is a very weak ground-state line of Fe II with a wavelength of 1901.773 \AA{} \citep{Morton2003}; therefore, we examined on-the-fly calibrated STIS spectra \citep{Micol} for the 14 lines of sight in this study and 3 lines of sight in SRF2002 where the data exist (including HD 24534, which we have reanalyzed).  When more than one observation exists for a given line of sight, we coadded all available data sets for the echelle order near 1900 \AA{} in the hope of detecting this weak Fe II feature.  We detect this line and measure its equivalent width in only two spectra (HD 24534 and HD 93222; see \S \ref{ss:1902line} for more details) and measure upper limits on the equivalent width in the other 16 spectra where we do not detect it.  To analyze this line, when detected, we assumed an empirically measured PSF (S. V. Penton, private communication) for the PSF of the E230H grating and the 0.2''$\times$0.09'' and 0.1''$\times$0.09'' apertures.  This empirical PSF is not Gaussian; however, the details of the PSF are unimportant, because measuring the equivalent width (as opposed to detailed velocity structure) of this very weak line should be fairly insensitive to the precise details of the PSF assumed.



\subsection{Observed Fe II Absorption Lines}
\label{ss:FeII_lines}
We follow the same procedure outlined in our previous work (\citeauthor{JensenOI}~\citeyear{JensenOI}, \citeyear{JensenNI}; SRF2002).  Specifically, we measure the equivalent widths of as many Fe II absorption lines in the {\it FUSE} wavelength region as possible, then fit the measured equivalent widths to a curve of growth with a single velocity dispersion (i.e.~$b$-value).  We use the seven Fe II absorption lines used by SRF2002 that are easily observed in most lines of sight.  We measure all seven of these absorption lines in nearly all the lines of sight in our sample; in all cases we have at least the three equivalent widths required to produce a solution and meaningful error for the two free parameters of column density $N$ and velocity dispersion $b$.  Other absorption lines exist in the {\it FUSE} wavelength region but are omitted from our study due to one or more of the following reasons:  (1) unknown $f$-value; (2) an $f$-value that implies the line is too weak to detect at typical column densities and {\it FUSE} S/N; or (3) interference from other atomic or molecular hydrogen lines.  One of these lines is worth mentioning specifically.  The line at 1121.97 \AA{} was used in the SB1979 and JSS1986 studies but is omitted from our study because (1) there could be small amounts of contamination from some weak neutral carbon lines and (2) its value of $f\lambda$ is very similar to that of the 1143.23 \AA{} line and, as the latter is rarely on the linear portion of the curve of growth, the former does not provide significant additional information.

Wavelengths and $f$-values for each of these lines used are summarized in Table \ref{linetable}.  The compilation of \citet{Morton2003} cites the theoretical values calculated by \citet{RU1998}.  However, summarizing previous results and using new {\it FUSE} data, \citet{Howk} empirically derived $f$-values for the Fe II lines used in this study and in SRF2002.  The concern with empirical $f$-values is that systematic errors will arise if the assumption of a single Gaussian velocity dispersion (i.e.~a single $b$-value), at least as an approximation, is not justified.  This is a concern especially with instruments that are not able to resolve fine velocity structure, such as the medium-resolution {\it FUSE}.  However, SRF2002 found that the \citet{Howk} $f$-values produced largely self-consistent single velocity dispersion curves of growth for 18 lines of sight; we find the same for our sample of 51 lines of sight.  We do not find this same consistency in test cases using the \citeauthor{RU1998} $f$-values, and have therefore elected to use the \citeauthor{Howk}~$f$-values in our analysis.  This choice is discussed in more detail in \S \ref{ss:f-values}.  Note that damping constants are not reported for these lines by \citet{Morton2003}, but that column densities are small enough that damping constants should not have a significant effect on our curves of growth.

\begin{figure}
\begin{center}
\epsscale{1.00}
\plotone{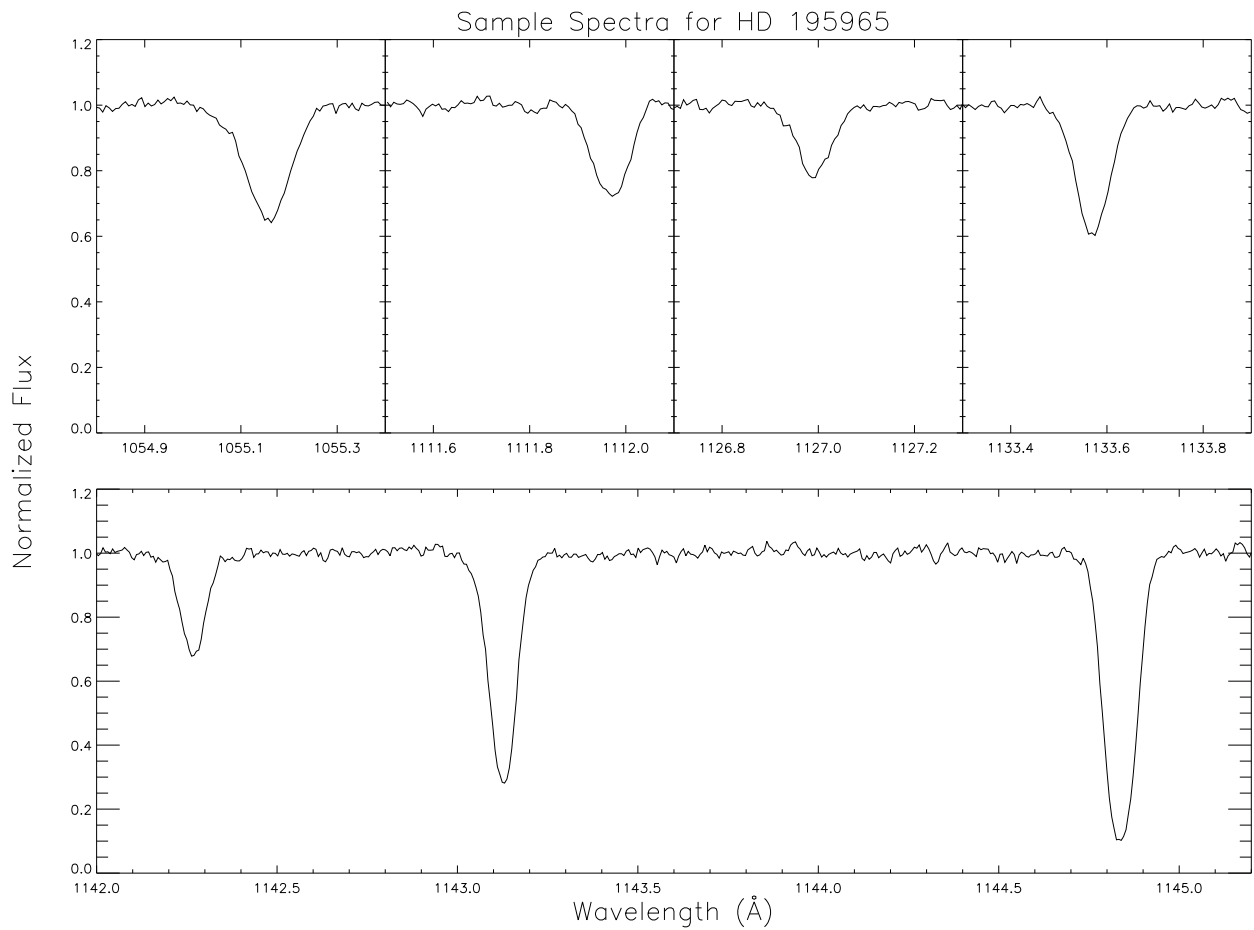}
\end{center}
\caption[Fe II Lines of HD 195965]{The seven absorption lines used in this study, shown for the HD 195965.  {\bf Top:}---From left to right, the 1054 \AA{} line, the 1112 \AA{} line, the 1126 \AA{} line, and the 1133 \AA{} line.  These four plots are shown with 0.6 \AA{} windows.  {\bf Bottom:}---The triplet of iron lines found between 1142--1145 \AA{}.  All spectra are shown with the local spectra normalized but without any velocity correction.  Note that there is very little evidence that the overall velocity structure deviates from being reasonably approximated by a single Gaussian component.}
\label{fig:spec195965}
\end{figure}

An example of all seven observed absorption lines is given in Figure \ref{fig:spec195965}, for the line of sight toward HD 195965.  The line profiles in this particular line of sight show virtually no evidence of multiple cloud components, and the weakest lines should be relatively free of saturation.

\subsection{Equivalent Width Measurements and Errors}
\label{ss:errors}
We measure equivalent widths ($\eqw$) of the observed absorption lines.  These measurements are reported in Table \ref{eqwidths}.  Lines are fit to either Gaussian or Voigt profiles as appropriate to the strength of the line.  Voigt profiles are fitted iteratively and convolved with the assumed instrumental PSF (a Gaussian of FWHM $\sim15\kmpers$).  The Voigt profile reduces to a PSF-like Gaussian with FWHM $\sim15\kmpers$ for small enough column density and $b$-value, and can therefore be used on both weak and saturated lines.  The recorded equivalent widths are the equivalent widths of the fits.  Occasionally, some marginally resolved velocity components are observed (or implied through asymmetric profiles).  In these cases we fit the multiple observed components simultaneously and record the total equivalent width of the feature.

When absorption lines are fitted to a Gaussian using IDL's {\it CURFIT} procedure, equivalent width errors are taken from standard error propagation of the curve-fitting routine and the functional form of a Gaussian.  These errors include continuum placement error (from photon noise) added in quadrature with the errors in the fit.  When absorption lines are fitted to a Voigt profile, errors are not clear due to the lack of a simple functional form.  In these cases we estimate 1-$\sigma$ errors by measuring the standard deviation in the noise of the surrounding continuum (essentially the reciprocal of the S/N) and summing over the wavelength range where the fit is a nontrivial amount below the continuum (usually $\sim95$\% of the continuum).  The sum is multiplied at each point by the relative depth of the fit to the absorption line; e.g.~the wavelength range over which a line core reaches a zero flux level does not make a contribution to the error.  In essence, this method estimates the error in a 1-$\sigma$ continuum misplacement integrated over the bulk of the observed width of the line.  In test cases where we compare this method to lines that can be fit by Gaussians, this method of error estimation yields errors that are of the same order but occasionally somewhat more conservative (up to $\sim50$\% larger).

Potential errors from the placement of the zero level of the spectrum are not formally accounted for; however, the {\it FUSE} spectra of the lines of sight in this study have several very wide (${\rm FWHM} >1$ \AA{}) $\Hmol$ bands that constrain errors in the placement of the zero level over wide ranges of the {\it FUSE} bandwidth \citep{Rachford2002}; in most cases, the zero level is uncertain to only a few percent of the continuum level (with results in a correspondingly small uncertainty in equivalent widths due to this effect).  This is much smaller than the formal errors resulting from factors such as continuum placement error due to photon noise, and is therefore disregarded.

Similarly, curvature of the surrounding continuum is small for most absorption lines in most lines of sight and is not formally included in the error.  The few cases with curvature that is severe enough to consider affect most strongly the line at 1144.9 \AA{}, which is strongly saturated, with the observed profile reaching a zero flux level in the core in most cases.  Therefore, in these cases the equivalent width that is derived is relatively insensitive to the curvature of the continuum.

\section{METHODS}
\label{s:methods}

\subsection{The Curve-of-Growth Method}
\label{ss:cogmethod}
We fitted the measured equivalent widths of the various Fe II absorption lines to a curve of growth with a single velocity dispersion.  The damping constant used to construct the curves of growth was similiar in order of magnitude to the summation of the Einstein $A$ coefficients for the triplet of iron lines at $\lambda \lambda \lambda$ 1142, 1143, 1144.  The effects of our choice of damping constant should be trivial; only the 1144 \AA{} line is significantly into the ``flat'' portion of the curve of growth, where the damping constant still has only a minimal effect.

We found the minimum $\chi^2$ with respect to the two parameters of column density $N$ and $b$-value, then used the extrema of the confidence ellipses to find the 1-$\sigma$ errors in $N$ and $b$-value.  These errors are calculated for a $\chi^2$ distribution of one degree of freedom, in essence holding either $N$ or $b$ fixed while determining the error on the other.  The best fits for these curves of growth are reported in Table \ref{coltable}, with 1-$\sigma$ errors.  Also reported in Table \ref{coltable} are the abundances of iron relative to hydrogen (values in the table are multiplied by $10^{7}$ for clarity).  The errors on these iron abundances are taken from standard error propagation of the iron and hydrogen column densities, i.e.~the errors are added in quadrature.

As is always the case with a curve-of-growth method, there are potential systematic errors stemming from our assumption that the velocity structure in each line of sight is reasonably approximated by a single Gaussian velocity dispersion ($b$-value).  The large $b$-values that we typically derive are indicative of several velocity components; in some cases, we see slightly resolved velocity structure.  However, we have reasons to believe that the velocity structure is reasonably approximated by a single $b$-value.  The lines we use span over 1.5 orders of magnitude in $f\lambda^2$, and our curves of growth are always very self-consistent, with at least some of the weaker lines being on the linear portion of the curve of growth.  Any combination of weak lines that show an increase in equivalent width proportional to $f\lambda^2$ confirms that there is little to no undetected saturation in those weak lines, as such a linear increase would not occur if a significant fraction of the total column density were found in saturated velocity components.  Therefore, we conclude that our curves of growth and the subsequently derived column densities and $b$-values are generally free of systematic errors arising from the single $b$-value assumption.  SRF2002 came to the same conclusion in their study.  Adopted curves of growth are shown in Figures \ref{fig:cogs1-15}-\ref{fig:cogs46-51}.

As a quantitative check on our conclusion that systematic errors are not significant, Table \ref{coltable} also lists the reduced $\chi^2$ values.  It can be seen that many of the reduced $\chi^2$ values are not $\leq1$, including a tail of values up to around $\sim10$.  There are at least three possible reasons for this.  First, artificially large reduced $\chi^2$ values may come from underestimated errors in the equivalent width measurements.  Our methods of estimating the error in the equivalent widths were described above; primarily, we assumed that continuum placement is the dominant source of error.  If this assumption is not valid, then our equivalent width errors on some lines may be too small.  We should note that we have also determined the best fits using ``unweighted'' curve-of-growth fits to determine if there are systematic effects due to our error estimation, e.g.~if small errors on certain lines unduly influence the curve-of-growth results.  The weighted and unweighted fits possess very close agreement; 47 of the 51 unweighted fits fall within the errors determined by the weighted calculations.  Therefore, we conclude that our fits are in fact the best results using a single velocity-dispersion approximation.

A second reason the reduced $\chi^2$ values may be larger than one is that there is a real deviation from a single velocity-dispersion curve of growth.  As discussed above, there is clear evidence that some of our lines of sight contain multiple cloud components, although we are still often able to achieve good fits.

The third probable reason for large reduced $\chi^2$ values regards the $f$-values that are used to calculate the curve-of-growth fits.  We discuss our choice of $f$-values below (\S \ref{ss:f-values}).  While adopting the $f$-values of \citet{Howk}, we also argue that the \citet{Howk} $f$-value for the 1112 \AA{} line may be in error.  If the 1112 \AA{} line is omitted, the reduced $\chi^2$ goes down significantly in many lines of sight; for example, from 7.2 to 1.7 for HD 195965 and 4.8 to 0.4 for HD 90087 (the derived column densities, however, remain similar).  To summarize, while some of the reduced $\chi^2$ values are larger than 1, we are still confident that the derived column densities are the best fits under the single velocity-dispersion assumption, and that this assumption is a reasonable approximation for most lines of sight.  The possibilities that equivalent width errors have been underestimated and/or that the 1112 \AA{} line's $f$-value is inaccurate could easily explain why the reduced $\chi^2$ is not less than 1 for some lines of sight.

A comparison of iron column densities in the four lines of sight that have been observed by both {\it FUSE} (either this paper or SRF2002) and \citet{Miller2007} shows some discrepancies.  The Fe II column density of HD 152590 in this paper agrees with the \citet{Miller2007} result within 1-$\sigma$, but HD 147888 does not, just barely missing agreement to within 2-$\sigma$.  The SRF2002 column density HD 207198 agrees with \citet{Miller2007} within 2-$\sigma$ but not quite 1-$\sigma$, while the two measurements of HD 27778 are approximately 3-$\sigma$ apart.  It is worth noting that while \citet{Miller2007} use the somewhat preferable method of fitting only weak lines, different weak lines are used for the different lines of sight.  The 2249 \AA{} line of Fe II \citep[with an $f$-value determined by][]{Howk} was used for the two lines of sight where the agreement is better, HD 152590 and HD 207198.  The 2260 \AA{} line was used for HD 27778, and the 2367 \AA{} line for HD 147888.  While we cannot comment quantitatively on potential systematic errors, it is clear that such errors may exist, particularly for these latter two lines of sight where the $f$-values used were from different sources.  In the case of the 2367 \AA{} line, only a theoretical $f$-value is available, while the other $f$-values are derived empirically from astrophysical data or in laboratory experiements.



\subsection{Hydrogen Column Densities}
\label{ss:FeII_hydrogen}
In order to examine abundances and depletions we need reliable measures of hydrogen.  We use values from the literature where they exist for both atomic and molecular hydrogen (with one exception for H I); in a few cases, we derived our own values.  When multiple values from the literature exist \citep[or, in the case of HD 202347, where we compare our derived $\NHI$ to a published value from][]{Andre2003}, we select the value with the smallest error.  These data are summarized in Table \ref{hydrogentable}.

The majority of atomic hydrogen measurements are published measurements of the Lyman-$\alpha$ profile of H I.  Only a few stars (HD 38087, HD 43384, HD 91597, HD 147888, HD 179406, and HD 210121) are cool enough that stellar contamination of the Lyman-$\alpha$ profile might be significant.  Atomic hydrogen in the line of sight toward HD 91597 was measured by \citet{DS1994}, who estimated the contribution from stellar Lyman-$\alpha$ through Str\"omgren photometry.  In Table \ref{hydrogentable} we record the interstellar value.  The hydrogen content of the HD 147888 line of sight was measured by \citet{Cartledge2004}, who used the hydrogen content of HD 147993 as a proxy.  However, this value is the same as that derived from the profile of the HD 147888 Lyman-$\alpha$ line, and is consistent within the errors with the value of \citet{FM1990}.  Thus we conclude that stellar contamination is insiginificant.  We made measurements of the Lyman-$\alpha$ line for seven lines of sight using STIS data.  The H I content of the HD 210121 line of sight was measured through 21 cm emission by \citet{WF1992}.

\citet{Bohlin1978} showed for less reddened lines of sight that there exists a correlation between selective reddening and total hydrogen column density such that $\NHtot=5.8\times10^{21}\ebv$.  \citet{Rachford2002} showed that this correlation remains valid to approximately $\pm0.30\dex$ (a factor of 2) in $\logHtot$ even for large hydrogen column densities ($\logHtot\gtrsim21$) and large molecular fractions of hydrogen; furthermore, the correlation is much more precise than $\pm0.30\dex$ in most cases.  We have assumed this correlation for HD 38087 and HD 43384, assuming that the $\pm0.30\dex$ error is in H I, and then have carried standard error propagation through for the total error in $\NHtot$.  For HD 179406, we quote the published value of \citet{Hanson1992}, who used a similar relationship between $\NHI$ and $\ebv$ from \citet{SvS1985}.

The molecular hydrogen measurements are determined for each line of sight by fitting several low-$J$ lines.  For details of this method see \citet{Rachford2002}.  The measurements are taken from several different sources (including the private communication of work in preparation and one measurement made by us for this paper, of HD 109399), but the techniques are largely the same.  A few of these lines of sight where we cite unpublished values for $\NHmol$ do have values of $\NHmol$ in the literature \citep[e.g.][]{Andre2003}, but we have elected to use the unpublished values on the basis of the smallest errors.

\section{RESULTS AND DISCUSSION}
\label{s:FeII_results}
As stated above, our derived column densities are given in Table \ref{coltable}, while curves of growth can be seen in Figures \ref{fig:cogs1-15}-\ref{fig:cogs46-51}.  For our sample (not including the unique lines of sight from SRF2002) we find that Fe II/$\Htot=3.2\pm0.1\times10^{-7}$, an average weighted by the inverse squares of the errors in the abundances.  This is somewhat smaller than the results of SB1979, Fe II/$\Htot=4.5\times10^{-7}$; however, the two numbers are calculated differently.  Our calculation is a weighted average of the abundances in each line of sight; the SB1979 abundance is a ``total'' abundance, calculated from the ratio of total observed Fe II to total observed hydrogen.  If we calculate a total abundance in this manner, we obtain a total abundance that is larger, Fe II/$\Htot=5.6\times10^{-7}$.  The weighted average tends to favor the smaller abundances which also have smaller errors, whereas calculating a total abundance slightly favors the tail of larger abundances, which are up to a few times larger than the median.

Whether we consider an average or the total iron abundance, our results are distinctly different from the results of SRF2002; the 16 lines of sight unique to that study have a weighted average of $1.9\pm0.2\times10^{-7}$ and a total abundance of $3.5\times10^{-7}$.  The fact that our abundances are larger in both cases has a simple interpretation---our study covers some of the same highly-reddened ground as SRF2002, but with an additional sample of much less-reddened lines of sight.  We note here that the averages of JSS1986, discussed in \S \ref{s:FeII_intro}, correspond to $1.4^{+0.2}_{-0.1}\times10^{-6}$ for the ``warm'' ISM and $3.5\pm0.5\times10^{-7}$ for the ``cold'' ISM; our weighted average is similar to the latter sample, but our total abundance falls squarely between the two (as it should if our sample covers both ranges of the two JSS1986 samples).

Much more insightful than the averages, however, are that we also see several statistically significant correlations between iron depletion and various line of sight parameters; these correlations and their interpretations are discussed in \S \ref{ss:FeII_correlations}.  In these plots and analyses, we include the results of SRF2002 but not SB1979 or JSS1986.  As discussed in \S \ref{s:FeII_intro}, our sample, compared to these previous samples, is either larger or covers more parameter space.  There are potential systematic errors between the {\it FUSE} studies (this paper and SRF2002) and SB1979 and JSS1986.  SB1979 used the 1122.0, 1133.7, and 1144.9 \AA{} lines of Fe II, while JSS1986 used the 1096.9, 1122.0, and 1133.7 \AA{} lines.  In addition to the differences in absorption line sets between those studies and ours, SB1979 used $f_{1133.7}=6.3\times10^{-3}$ and $f_{1144.9}=0.15$, while JSS1986 adopted $f_{1133.7}=4.8\times10^{-3}$.  As noted in Table \ref{linetable}, we used the $f$-values adopted by \citet{Howk} of $f_{1133.7}=5.5\times10^{-3}$ and $f_{1144.9}=0.106$.  However, in comparing their results with SB1979, SRF2002 concluded that the SB1979 abundances should be increased (due to the downward revision of $f$-values) by an average of 20\%, much smaller than the inherent variation in abundances.  Therefore, there is not a clear reason to suspect that $f$-value differences lead to significant systematic errors.  On the other hand, the fact that this study and SRF2002 are based on more absorption lines (7 in most cases) covering a broader range in $f$-value than the studies by SB1979 and JSS1986 (3 lines at most) is one reason to prefer the {\it FUSE} results; at the least, it justifies including only the SRF2002 results in our analysis.

Our search for the weak 1901.773 \AA{} line of Fe II and the two probable detections that resulted are discussed in \S \ref{ss:1902line}.  A brief comparison of the previous theoretical \citep{RU1998} and empirical \citep{Howk} $f$-values is given in \S \ref{ss:f-values}.  Finally, we discuss the overall Galactic abundance of iron (and, briefly, the implications for dust models) in \S \ref{ss:cosmic_iron}.

\subsection{Correlations}
\label{ss:FeII_correlations}
In this subsection we discuss the correlations and anticorrelations found between iron abundances and depletions and various line of sight parameters (measures of hydrogen content, extinction, reddening, and line of sight pathlength).  As discussed, above, we have included the lines of sight from SRF2002.  We use the Fe II column densities determined in that paper, but in order to calculate abundances and depletions and analyze correlations we find our own values from the literature for the line of sight parameters just mentioned.  This is the reason for any observed discrepancies between our plots and the plots in SRF2002.  We have not included the data points from either SB1979 or JSS1986 for the reasons discussed above.

Correlations between iron depletion and measures of hydrogen (total hydrogen column density, average hydrogen volume density, and the molecular fraction of hydrogen) are discussed in \S \ref{sss:FeII_hydro_corr}; correlations between iron depletion and extinction and reddening parameters are discussed in \S \ref{sss:FeII_ext_corr}; an anticorrelation between iron depletion and distance (i.e.~line of sight pathlength) is discussed in \S \ref{sss:FeII_dist_anticorr}; and lastly, we provide a few comments on outlying data points in \S \ref{sss:outliers}.

\subsubsection{Correlations with Hydrogen}
\label{sss:FeII_hydro_corr}
Iron depletion shows a clear correlation with increased average hydrogen volume density; see Figure \ref{fig:logFeIIHlognh}.  This is not surprising; this trend was already known for Fe II (JSS1986, SRF2002), and known or at least suggested for several other elements (JSS1986; \citeauthor{Cartledge2001}~\citeyear{Cartledge2001}, \citeyear{Cartledge2004}).  The physical explanation usually invoked for this phenomenon is the model by \citet{Spitzer1985}, discussed in \S \ref{s:FeII_intro}, that measurements of individual lines of sight are probing two fundamental cloud types, warm and cold, and that the relative mix of the two cloud types along the line of sight determines the observed depletions.  The average hydrogen volume density, $\nHavg$, is usually taken to be the best diagnostic for the determining the mix of the two cloud types being probed.

In general, we have not probed significantly larger $\nHavg$ than previous studies, with one exception:  HD 147888 has an average hydrogen volume density of 13.6 cm$^{-3}$, more than twice as dense than any of the lines of sight in this sample, SRF2002, or SB1979.  (Note that HD 179406 is also more dense than any line of sight in SRF2002, though only half as dense as HD 147888.)  While 147888 is among the most highly depleted lines of sight in this sample ($\logFeIIH=-6.87$), it does not show depletion that is particularly extreme; seven other lines of sight have $\logFeIIH<-6.8$ within their errors.  Although we do not want to overstate the importance of one data point, this line of sight could potentially support the \citet{Spitzer1985} model at least up to this larger $\nHavg$.  It would seem that even in this dense line of sight we are not probing clouds with the physical conditions expected of translucent clouds---or, perhaps, models of translucent clouds need to be adjusted to explain the observed lack of extreme depletions.

We also find that iron depletion is correlated with the total hydrogen column density.  \citet{WakkerMathis} have suggested that $\NHI$ is in and of itself a good indicator of depletions.  While the correlations of depletion with respect to both $\NHI$ and $\nHavg$ are similar, we examined the partial correlation coefficient of the iron abundance and hydrogen volume density while hydrogen column density held constant (an example of using partial correlation coefficients in this manner is outlined in JSS1986).  When we examine the correlation of the logarithmic iron abundance with the logarithm of $\nHavg$ and $\logHtot$, the partial correlation coefficient of the iron abundance and $\nHavg$ is -0.447.  Using a $t$-distribution for the appropriate number of degrees of freedom (in the manner of JSS1986), we find that the probability that the observed correlation coefficient has an absolute magnitude this large if the true correlation coefficient is 0 (we will refer to this as the probability of the null hypothesis) is 0.03\%.  If we examine the converse, the partial correlation between iron depletion and $\logHtot$ with $\lognHavg$ held fixed is still significant: the coefficient is -0.267; the two-sided probability of the null hypothesis is 3.4\%.  These results imply that while observationally speaking, both $\lognHavg$ and $\logHtot$ are very good predictors of abundances and depletions, $\nHavg$ appears to have more fundamental physical significance.  We note here that $\logFeIIH$, $\lognHavg$, and $\logHtot$ were chosen as opposed to their linear counterparts because these variables show the strongest correlations.  We also note that using these correlation coefficients makes several implicit assumptions, including a linear correlation between the variables being examined, and Gaussian distributions of the variables for the true overall populations.  The validity of these assumptions potentially influences the validity of the above analysis, but we do conclude that $\nHavg$ is more significant.


SB1979 noted a possible rough correlation between iron depletion and the molecular fraction of hydrogen, $\fHmol$.  However, SRF2002, including the SB1979 points, did not conclude that there was an overall correlation.  Examining Fig. 5 of SRF2002 reveals that in general, the lines of sight with a large molecular fraction (primarily the {\it FUSE} lines of sight) have relatively constant depletion, while lines of sight with a smaller molecular fraction (primarily the {\it Copernicus} lines of sight) show a great deal of scatter in iron depletion (over an order of magnitude in the iron abundance).

Analyzing only our data points and those from SRF2002, we {\it do} observe an overall correlation between iron depletion and $\fHmol$; see Figure \ref{fig:logFeIIHHf}.  With the exception of a few lines of sight, this is in fact one of the strongest correlations seen in our data.  We have briefly explored the possibility that in regions with a large $\fHmol$ this additional iron is found in the form of gas-phase Fe I, but conclude that the gas-phase Fe I abundance is, in general, at least an order of magnitude too small to account for the difference.


Despite the strength of the overall correlation, there are several outlying points.  Lines of sight with a large $\fHmol$ but little depletion include HD 210121 (from our sample) and HD 27778 and HD 62542 (from SRF2002), while lines of sight with a small $\fHmol$ but larger depletions include HD 147888, HD 152236, and HD 164740.  We will return to these outlying data points in \S \ref{sss:outliers}.  Ultimately we conclude that while $\fHmol$ is strongly correlated with iron depletion, it is not a perfect indicator of the physical conditions required for large depletions, and environments that are merely very dense without necessarily forming significant $\Hmol$ can still show large depletions.  As above, to examine the independent significance of the $\fHmol$ correlation, we examined the partial correlation coefficients between iron depletion and $\fHmol$ with $\nHavg$ held fixed, we conclude that the correlation with $\fHmol$ is largely secondary to the correlation with $\nHavg$.  If the three variables considered are $\logFeIIH$, $\fHmol$, and $\lognHavg$, the probability of the null hypothesis is well over 50\% whether $\fHmol$ is considered linearly or logarithmically.  If the linear values of the iron abundance or $\nHavg$ are used, the probability of the null hypothesis $\nHavg$ decreases (i.e.~that the correlation is more real is likely); however, as before, $\logFeIIH$ and $\lognHavg$ are examined because these two variables exhibit a much stronger correlation than the linear variables.  Again, we note the implicit assumption of linear correlations between whatever variables (linear, logarithmic, or otherwise) are under examination.

In the sense that the correlation with $\fHmol$ is not necessarily independently significant, we have not shown anything new.  However, we have shown that in our sample a stronger correlation exists than in previously studied samples (e.g.~SB1979, JSS1986, SRF2002).  Why the SB1979 sample shows such a large spread in depletion at lower values of $\fHmol$ that is not seen in our sample is unclear.  We have discussed earlier the potential for systematic errors between {\it Copernicus} and {\it FUSE} studies, but even if those errors are larger than we estimate, they do not account for the spread within the self-consistent {\it Copernicus} data.  Scattered light, which was worse for {\it Copernicus} than {\it FUSE}, may be a small factor in causing some variations, but is almost certainly not responsible for the majority of the scatter.  Inherent variation over the short pathlengths of some of the SB1979 targets may also be factor.  It is worth noting that while our results show a few points of scatter at both large and small $\fHmol$, the SB1979 scatter is exclusively at small $\fHmol$.  At $\fHmol \gtrsim 0.3-0.4$, there are no SB1979 targets with small depletions, but rather the few SB1979 targets with $\fHmol$ this large more or less conform to the trend seen in our sample.  This compares well with \citet{Cartledge2006}, where trends of depletion for elements such as magnesium are also seen with respect to $\fHmol$---though not as strong as between depletion and $\nHavg$---and with some scatter at small $\fHmol$.



\subsubsection{Correlations with Extinction and Reddening Parameters}
\label{sss:FeII_ext_corr}
Figure \ref{fig:logFeIIHred} shows $\logFeIIH$ as a function of four different reddening or extinction parameters---the relative color excess between the $B$ and $V$ bands, $\ebv$; $\ebv$ scaled by line-of-sight pathlength, $\ebvdist$ the total visual extinction, $\av$; and $\av$ scaled by line-of-sight pathlength, $\avdist$.  Both $\ebv$ and $\av$ are measured in magnitudes, and we calculate $\ebvdist$ and $\avdist$ in magnitudes per parsec.

The parameter of $\ebv$ is known to correlate well with $\NHtot$, and is thought to be strongly correlated with the total dust column density in the line of sight.  Given the correlation we see between iron depletion and $\NHtot$, it is not surprising that we also see a correlation between iron depletion and $\ebv$---denser environments are correlated with both iron depletion and total dust content.  SRF2002 noted that the correlation between iron depletion and $\ebv$ seems to break off at $\ebv \approx 0.35 \magnitude$.  We do not observe this trend.  In our combined sample, we see a very strong correlation overall.  The slope we see for the entire sample is about half as steep as the slope seen in SRF2002 for lines of sight with $\ebv < 0.35$, but a few times steeper than the slope of the lines of sight with $\ebv > 0.35$ in that paper.  If we take the same approach as SRF2002 and look for the point where the correlation disappears for larger $\ebv$, we must restrict the sample to the 23 lines of sight with $\ebv \geq 0.5\magnitude$.  In this sense, we still see the same effect noted in SRF2002, which they interpreted as a threshold effect, wherein the conditions that increase $\ebv$ do not necessarily require additional iron per hydrogen atom.

The parameter $\av$ also correlates well with the total hydrogen column density and should also be correlated with the total dust column density.  As discussed in \S \ref{s:FeII_intro}, SRF2002 did not find any correlation between iron depletion and $\av$ for the 18 lines of sight in that paper.  They were not able to compare their 18 lines of sight with the SB1979 or JSS1986 results because those papers did not report on $\av$ for those lines of sight.  However, taking our sample and the SRF2002 sample, we do find a correlation between iron depletion and $\av$.  The reason for this difference seems to be the wider range of $\av$ examined by our sample.  When we restrict the sample to lines of sight with $\av \geq 1.5\magnitude$, the correlation weakens, and it disappears for lines of sight with $\av \geq 2\magnitude$ (at this point, however, it should be noted that the restricted sample contains only 11 lines of sight).

If $\ebv$ and $\av$ are correlated with the total dust column density, then dividing by the pathlength of the line of sight results in quantities ($\ebvdist$ and $\avdist$) that should be correlated with the total dust volume density.  Not surprisingly, iron depletion is correlated with both of these quantities, at least as strongly as it is correlated with the corresponding extinction parameters integrated over the entire line of sight.  Calculating partial correlation coefficients indicates that the probability of the null hypothesis for a correlation between iron depletion and $\ebvdist$ with $\nHavg$ held constant is between 10\% and 20\%, for all possible combinations of linear and logarithmic quantities.  The relationship between the iron abundance and $\avdist$ with $\nHavg$ held constant is less clear; the probability of the null hypothesis is 42\% if $\avdist$ and $\nHavg$ are logarithmic, but less than 1\% if they are linear.  Reversing the correlations, we see that iron depletion is always very well-correlated with $\nHavg$ even when other variables are held constant.  We therefore conclude that the most important variable of interest for determining iron depletion is still $\nHavg$, but that these measures of dust density ($\ebvdist$ and $\avdist$) are also useful quantities that independently correlate with the iron depletion to a small degree.  That the measures of dust density should correlate with iron depletion is also physically motivated in that the depleted material should be found in the dust.  That $\nHavg$ is still the variable most strongly correlated with iron depletion potentially indicates that the way iron is incorporated into the dust is nonuniform, and/or that the dust grains that contain some iron have variable extinction characteristics.


Dividing these extinction parameters by line of sight pathlength also somewhat resolves the ``threshold'' effects seen for the integrated extinction parameters alone.  Correlations, though both weaker and less significant, remain---rather than disappear---even when only the densest lines of sight are considered.  However, the fact that the correlations do weaken and we do not see extreme depletions indicates that we are not observing lines of sight that are dominated by translucent clouds, though perhaps some lines of sight do probe these clouds to a limited extent.


The quantity $\rv$ is the ratio of total visual extinction to selective extinction ($\rv$ is defined as $\av / \ebv$) and is correlated with grain size (because larger grains contribute significantly to $\av$ but not to $\ebv$).  We find that while there is the hint of a correlation between iron depletion and with $\rv$ (e.g.~a linear regression of the linear abundance and $\rv$ has a statistically significant slope), upon further examination the correlation is not particularly significant.  For example, using a Pearson correlation coefficient, the probability of the null hypothesis for logarithmic iron depletion and $\rv$ is 35\%.  Rank correlation coefficients (Spearman's $\rho$ and Kendall $\tau$, using IDL's R\_CORRELATE function) that do not carry any implicit assumptions about the functional form of the correlation also show that any correlation is slight at best.  As discussed below in \S \ref{sss:outliers}, we also find anecdotal cases where lines of sight that do not conform to some of the observed trends (e.g.~$\fHmol$) tend to be correlated with $\rv$ values different from the interstellar average of 3.1; however, we cannot draw the conclusion that there is an overall correlation between depletion and $\rv$.

Lastly, we will point out a few interesting anecdotal cases regarding cases of extreme depletion and large reddening and extinction.  Given that the overall trends hold, we could point out many cases, but we will restrict ourselves to the three cases with the largest depletions:  from most to least depletion, HD 164740, HD 110432, and HD 147888.  When considering the combined sample, HD 164740 has the largest $\av$ and second largest $\ebv$, HD 110432 has the second largest $\avdist$, and HD 147888 has the largest $\avdist$ and $\ebvdist$.  And though we do not claim a conclusive trend between depletion and $\rv$ overall, we note that these three lines of sight all have large $\rv$ of $\gtrsim4$, including HD 164740 with the second largest $\rv$ in the sample of 5.36.

\subsubsection{Anticorrelation with Distance}
\label{sss:FeII_dist_anticorr}
We find that iron depletion is anticorrelated with distance, i.e.~line of sight pathlength.  With the exception of a few outlying points (HD 210121 from our sample and HD 27778 and HD 62542 from SRF2002), iron depletion decreases out to line of sight pathlengths of about 2 kpc.  For lines of sight with pathlengths from about 2--6 kpc, iron depletion appears constant to within $\sim0.4\dex$, significantly less variation than the decrease of an order of magnitude found in lines of sight with pathlengths less than 2 kpc (again, ignoring the three outlying points).  

The fact that we see constant depletion for long pathlengths is sensible for two reasons.  First, lines of sight with pathlengths of at least a few kiloparsecs should be sampling the conditions of several clouds, and therefore the total iron abundances and depletions should remain relatively constant.  Secondly, these longer lines of sight are, on average, less dense and less reddened per unit distance than the shorter lines of sight.  However, it cannot be assumed that the lines of sight are truly uniform; slight variations, therefore, may nevertheless contribute to our observed correlations with other line of sight parameters (such as reddening and extinction).



\subsubsection{Outlying Data Points}
\label{sss:outliers}
We have already mentioned six outlying data points.  HD 27778, HD 62542, and HD 210121 have smaller iron depletions but larger values of $\nHavg$, $\fHmol$, $\ebvdist$, and $\avdist$, breaking the trends that we have just discussed.  Conversely, HD 147888, HD 152236, and HD 164740 have large iron depletions despite small values of $\fHmol$.

We will first approach the somewhat easier task of interpreting the second set of outliers.  HD 147888, HD 152236, and HD 164740 all have very large hydrogen column and/or volume densities.  Additionally, these lines of sight all have relatively large values of extinction and reddening---HD 147888 has the largest $\avdist$ and $\ebvdist$ in the sample, while HD 152236 and HD 164740 both have large pathlength-integrated values of $\av$ and $\ebv$ (when scaled by pathlength, HD 152236 is in the upper half of the sample for both parameters and HD 164740 is in the upper third).  Therefore, given the trends we see with respect to all of these parameters, it is not surprising that we see large iron depletions.

To explain the low fractions of $\Hmol$, we note that all three of these lines of sight have $\rv$ larger, and in two cases much larger, than the interstellar average of 3.1:  4.06 for HD 147888, 3.29 for HD 152236, and 5.36 for HD 164740.  Large values of $\rv$ imply a large average grain size, but a large average grain size also implies that the $\Hmol$ formation rate may be small because of the reduced surface area per unit volume \citep[see][]{Snow1983}.  Comparing the three lines of sight, HD 152236 has the smallest iron depletion, smallest $\rv$, and largest $\fHmol$, while HD 164740 has the greatest iron depletion, largest $\rv$, and smallest $\fHmol$, with HD 147888 in the middle in all cases.  These trends fit our interpretation that these lines of sight have smaller molecular fractions of hydrogen due to increased grain size, but otherwise follow the trends between iron depletion and overall gas and dust density.  Additionally, we should note that the increased grain size is unlikely to cause the additional iron depletion, because the reduced average grain size that suppresses $\Hmol$ formation should also reduce the sticking rate of atoms and ions to grains.  Rather, the increased grain size is likely a result of grain coagulation in environments where iron is already highly depleted.

It is more difficult to interpret the other three lines of sight with relatively low iron depletions despite having large hydrogen volume densities, large molecular fractions of hydrogen, and larger values of reddening and extinction (both per unit length and pathlength integrated).  We first note that all three of these lines of sight have relatively large errors in the iron abundance, so it is possible that these lines of sight are merely statistical fluctuations in the overall correlation.  However, assuming this is not the case, there are a few comments that we can make about these lines of sight.

The atomic hydrogen column density that we report for HD 210121 is based on a 21 cm measurement from \citet{WF1992}.  If this measurement is saturated, then both the iron abundance and the molecular fraction of hydrogen are somewhat smaller and can be partially reconciled with the correlation.  We note that while the total hydrogen column density derived from adding the $\Hmol$ measurement of \citet{Rachford2002} and the $\NHI$ measurement of \citet{WF1992} implies $\logHtot=21.00$, the relationship of \citet{Bohlin1978}, which correlates $\ebv$ with $\Htot$, implies that $\logHtot=21.37$.  Also worth noting is that if the HD 210121 curve of growth is fit with the equivalent width measurements being artificially weighted equally (i.e.~an unweighted curve-of-growth fit), the result for the iron column density is substantially smaller (by $0.31\dex$).  Therefore, either an underestimated value of $\NHtot$ or an overestimate iron column density could be responsible for an iron abundance that seemingly deviates from certain trends.  It is worth noting that the difference between the weighted and unweighted fits is a phenomenon that is generally not observed in other lines of sight.

We have used the Fe II column density for HD 27778 from SRF2002, but we note that \citet{Miller2007} derived a much smaller Fe II column density (and resulting iron abundance) for this line of sight.  \citet{Miller2007} used the 2260 \AA{} line to derive the Fe II column density in HD 27778; we do not have an adequate basis for comparing the use of this line (primarily in terms of its $f$-value) to our curve-of-growth results.  However, we note than an adoption of the \citet{Miller2007} column density would eliminate this line of sight from consideration as an ``outlier.''

It is worth noting that the abundances of molecules such C$_2$ and CN are relatively large in HD 27778 \citep{Federman1994}, HD 62542 \citep{Gredel1993}, and HD 210121 \citep{Gredel1992}.  The properties of the diffuse interstellar band (DIB) absorption features are somewhat unique in the HD 62542 \citep{Snow2002, ABM2005} and HD 210121 \citet{Thorburn2003} lines of sight, with unusually strong ``C$_2$ DIBs'' (DIBs correlated with C$_2$) and unusually weak ``classical'' DIBs (other DIBs not correlated with C$_2$).  All three lines of sight also have $\rv$ at least slightly less than the interstellar average of 3.1: 2.73 for HD 27778, 2.83 for HD 62542, and 2.07 for HD 210121.

The last point, about the small values of $\rv$, implies that the average grain size is smaller in these lines of sight.  With a population of small grains, the surface area per unit volume is increased and the rate of $\Hmol$ formation may increase.  This may be the case for these lines of sight---supported by the fact that of the three lines of sight, HD 210121 has the smallest value of $\rv$, and presumably the smallest average grain size, but the largest fraction of $\Hmol$.  Also potentially supporting this interpretation are the large abundances of other molecules in these lines of sight.  However, an important caveat is that the increased surface area per unit volume should also increase sticking of gas-phase atoms to grains.  Therefore, it is still unclear why the iron depletions are so low.  If the smaller average grain size is the result of destruction processes (e.g.~shocks), then the answer may be that the destruction processes may somehow preferentially destroy iron-bearing grains, releasing iron back to the gas phase.  Again, however, it is very unclear why this would be the case, as shocks are thought to destroy grain mantles rather than grain cores, and the major source of depletion for iron should be grain cores, not mantles.

While it is beyond the scope of this paper to explore this issue in detail, we have also examined correlations between $\rv$, $\nHavg$, and $\fHmol$ in our entire sample.  We see, for instance, that $\fHmol$ increases with $\nHavg$ up to about $\nHavg \approx 2$ cm$^{-3}$, where the correlation begins to break down and, in fact, becomes an anticorrelation (though not as statistically significant as the correlation at low densities).  Therefore, we are possibly seeing evidence of the low $\Hmol$ formation rate in dense environments.  There is a significant amount of scatter in our attempt to correlate $\rv$ and $\fHmol$, with most lines of sight centered near $\rv=3.1$ with a wide range of $\fHmol$.  Nevertheless, there is an anticorrelation between $\rv$ and $\fHmol$, significant at the 1-$\sigma$ level, that can be found using several different correlation methods---a simple linear regression, in addition to Pearson, Spearman's $\rho$, and Kendall $\tau$ correlation coefficients.  This anticorrelation exists for both the entire range in $\rv$ and for a narrower range in $\rv$ near the ISM average of 3.1.  Again, given that $\rv$ is correlated with grain size, this could be interpreted as limited evidence that the $\Hmol$ formation rate is lower in environments with larger grains.  However, radiation may also be an important factor; this is briefly discussed below.  While less relevant to the issue of $\Hmol$ formation, we also note that there is not a statistically significant correlation between $\nHavg$ and $\rv$.

However, in spite of the explanation that variations in the correlation between the iron abundance and $\fHmol$ might be explained by unusual values of $\rv$, we should also note that unusual values of $\rv$ do not guarantee a break from this correlation.  For example, HD 38087, with the largest $\rv$ in our sample, does not break from the trend.  This may be in part due to the fact that while $\rv$ gives a rough measure of average grain size, it does not give precise information about the details a grain population, nor does it measure total grain surface area (as opposed to surface area per unit grain volume).  For example, while grain coagulation reduces the surface area per unit volume, an increased average grain size due to mantle growth will increase the total grain surface area \citep{Snow1983}.  An additional consideration is $\Hmol$-dissociating radiation.  The HD 147888 line of sight passes through the $\rho$ Oph cloud complex (HD 147888 is $\rho$ Oph D), which is known to have a high internal ultraviolet radiation field.  This may be generally true regarding lines of sight with large values of $\rv$---for example, Figure 1 of \citet{Draine2003}, calculated using the data of \citet{Fitzpatrick1999}, shows that for the same value of $\av$, a typical line of sight with $\rv=4$ has significantly less extinction (about 3 magnitudes) at 1000 \AA{} than a typical line of sight with $\rv=3.1$.  Thus, a higher far-UV radiation field allowed by the reduced far-UV extinction almost certainly plays some role in the small values of $\Hmol$ in some lines of sight with large values of $\rv$.  In some cases, the far-UV radiation may be the dominant cause.  However, it should be noted that \citet{Snow1983} nevertheless argues, based on the large number of lines of sight that are both dense and have small values of $\fHmol$ in the survey of \citet{ChaffeeWhite}, that a high radiation field is unlikely to be exclusively responsible for these cases.

In conclusion regarding the issue of grain size affecting $\Hmol$ formation rates, we have shown a few interesting anecdotal cases wherein applying this hypothesis to these outlying points seems to have some explanatory power.  However, this cannot necessarily be considered a uniform effect or the dominant cause of these outlying points.  Making definitive predictions would require further information about the grain population (namely, total grain mass) and the local radiation field.

\subsection{The 1901.773 \AA{} Line of Fe II}
\label{ss:1902line}
As mentioned in \S \ref{ss:FeII_HSTdata}, Fe II has a very weak transition at 1901.773 \AA{}.  This wavelength region is within the STIS wavelength coverage for 17 lines of sight in the combined sample of this paper and SRF2002 (near the edge of coverage in many lines of sight).  Prior to this paper, there have been no published detections of this line in interstellar absorption due to its small $f$-value \citep[calculated to be $7.00\times10^{-5}$ by][]{RU1998}, although weaker lines of Fe II at 2267 \AA{} \citep[$f=2.16\times10^{-5}$;][]{CardelliSavage1995} and 2234 \AA{} \citep[$f=1.29\times10^{-5}$;][]{Miller2007} have been detected.  We examined the available STIS data (see \S \ref{ss:FeII_HSTdata} for our methods) in an attempt to detect this feature, and our search resulted in two detections (HD 24534 and HD 93222) and 15 upper limits.  In this section, the detections will be discussed.  Below (\S \ref{ss:f-values}) we will discuss the upper limits and the derivation of an $f$-value from these data.  The results are recorded in Table \ref{eqwidths1902}.

In the spectrum of HD 24534, we find a small feature at 1901.85 \AA{} with a central depth of $\sim5\%$ of the continuum that we identify as the 1901.773 \AA{} line.  We cite the following pieces of evidence in favor of this identification:  (1) the velocity offset of $\approx13\kmpers$  precisely matches that of the dominant velocity component of other lines of Fe II and other elements seen in this line of sight, in particular the 1355.5977 \AA{} O I and the 1608.4511 line of Fe II (the latter of which is too saturated and too similar in $f$-value to the 1144.9 \AA{} line to be of additional use in this study); (2) the equivalent width is roughly consistent with the column density measurement, as discussed below; and (3) no other significant ground-state transitions of any abundant elements exist within 1 \AA{} of the Fe II line.

HD 24534 \AA{} was studied in SRF2002, who found a Fe II column density of $\logFeII=14.42^{+0.14}_{-0.13}$ and a $b$-value of $7.5^{+3.3}_{-2.0}\kmpers$.  Because we detected a feature that appears to be the 1901.773 \AA{} line of Fe II, we have reanalyzed the {\it FUSE} spectra to independently determine our own column density and $b$-value, $\logFeII=14.63\pm0.06$ and $b=6.4\pm0.5\kmpers$.  Though the difference in the column density of $0.21\dex$ is nontrivial, the column densities are nearly consistent within the 1-$\sigma$ errors.  The difference likely arises due to the fact that, because of our coadding procedures, we were able to detect the 1127 \AA{} line of Fe II (the weakest of the {\it FUSE} lines), which SRF2002 did not include in their analysis of HD 24534, and our equivalent widths are better constrained.

Based on the revised column density that we have derived and the \citet{RU1998} $f$-value of $7.00\times10^{-5}$, the expected $\eqw$ of the 1901.773 \AA{} line for HD 24534 is 0.96 m\AA{}.  This is in excellent agreement with our measured equivalent width of $0.94\pm0.17$ m\AA{}.  This fit, however, is subject to a possible systematic error.  The area where we identify the feature is slightly asymmetric, and the fit of the line depends on whether we choose a narrow fit that focuses on the portion of the feature where the depth is the greatest or a broader fit that covers the entire feature, including the redward asymmetry.  If we choose the latter fit, we obtain a larger $\eqw$ of $1.73\pm0.26$ (consistent with the other fit to within 2-$\sigma$).  Both fits have a similar reduced $\chi^2$ of $\approx 0.73$.  We have selected the narrower fit because of its consistency with the column density, but note that this choice ignores this possible systematic error.

We observe a similar feature in the spectrum of HD 93222 that we identify as this line.  In this spectrum (with a lower S/N than HD 24534), the feature is at a central wavelength of 1901.62 \AA{} and has a depth of $\approx10\%$ of the continuum.  For the same three reasons as before (matching of velocity offset with other lines, rough consistency with derived column density, and lack of other potential identifications), we identify this feature as the 1901.773 \AA{} line.

The expected $\eqw$ for HD 93222, based on the column density derived through the {\it FUSE} lines, is 5.76 m\AA{}.  However, in HD 93222 we identify at least two marginally resolved cloud components in the FUV absorption lines; the velocity offset of the feature we detect matches the velocity offset of the weaker, narrower component.  However, we have also analyzed the doublet of Mg II found at 1240 \AA{}, and in that case find three major components---at -23, -6, and 7$\kmpers$.  The feature we detect and identify as the 1901.773 \AA{} line is at -23$\kmpers$ with respect to its rest wavelength; the -23$\kmpers$ component is the strongest (i.e.~largest column density) component for Mg II.  If we assume that the velocity structure of Mg II and Fe II is similar, then the stronger Fe II component (at $\sim0\kmpers$) in the FUV lines is not a single stronger component but a blend of the two weaker components seen for Mg II (-6 and 7$\kmpers$), and the weaker Fe II component is identified with the strongest Mg II component (both at -23$\kmpers$).
We determined that the -23$\kmpers$ component contains approximately 43\% of the total column Mg II density.  Assuming the same distribution for Fe II implies an expected equivalent width of 2.45 m\AA{} for the -23$\kmpers$ component, which agrees within the errors with our measured value of $1.95\pm0.76$ m\AA{}.  Our simple simulations show that the smaller components (at -6 and 7$\kmpers$) of the 1901.773 \AA{} line may be simply lost in the noise.  

Both of these comparisons are based on the assumption that the \citet{RU1998} $f$-value is correct.  Working the above problem in reverse, we have taken our equivalent widths and column densities and calculated an empirical $f$-value in \S \ref{ss:f-values}.  Although the value we calculate is consistent with the \citet{RU1998} value, this line is worth more study to further constrain its $f$-value.  In any case, these detections at the very least roughly confirm the line's $f$-value, which in turn is extremely important for future work with COS, now scheduled to be installed on {\it HST} in 2008.  COS will have greatly increased sensitivity (depending on wavelength) but somewhat lower resolution than the highest dispersion modes of STIS.  It is possible that this line will be able to reveal Fe II column densities down to at least $\logFeII\approx14$ and possibly smaller, without fear of saturation effects.  Detecting such small column densities with COS may not be necessary, however, as many COS targets will have much larger total hydrogen column densities; thus, iron column densities will also be larger unless there are enhanced depletion effects.  If enhanced depletion effects are observed, however, this line may be of utility as our rough measure of its $f$-value is larger than the $f$-value of the weak Fe II lines used by \citet{Miller2007}.  Therefore, this line may be more easily detected in lines of sight with extreme depletion and/or poor S/N, while still being weak enough to guarantee reasonable freedom from saturation.

\subsection{Fe II $f$-values}
\label{ss:f-values}
\citet{Morton2003}, which is a definitive compilation of $f$-values and other atomic data for UV lines of astrophysical interest, quotes the $f$-values theoretically calculated by \citet{RU1998} rather than the more recent empirical $f$-values derived by \citet{Howk} using {\it FUSE} data.  However, SRF2002 used the \citeauthor{Howk}~$f$-values in their study.

\begin{figure}
\begin{center}
\epsscale{1.00}
\plotone{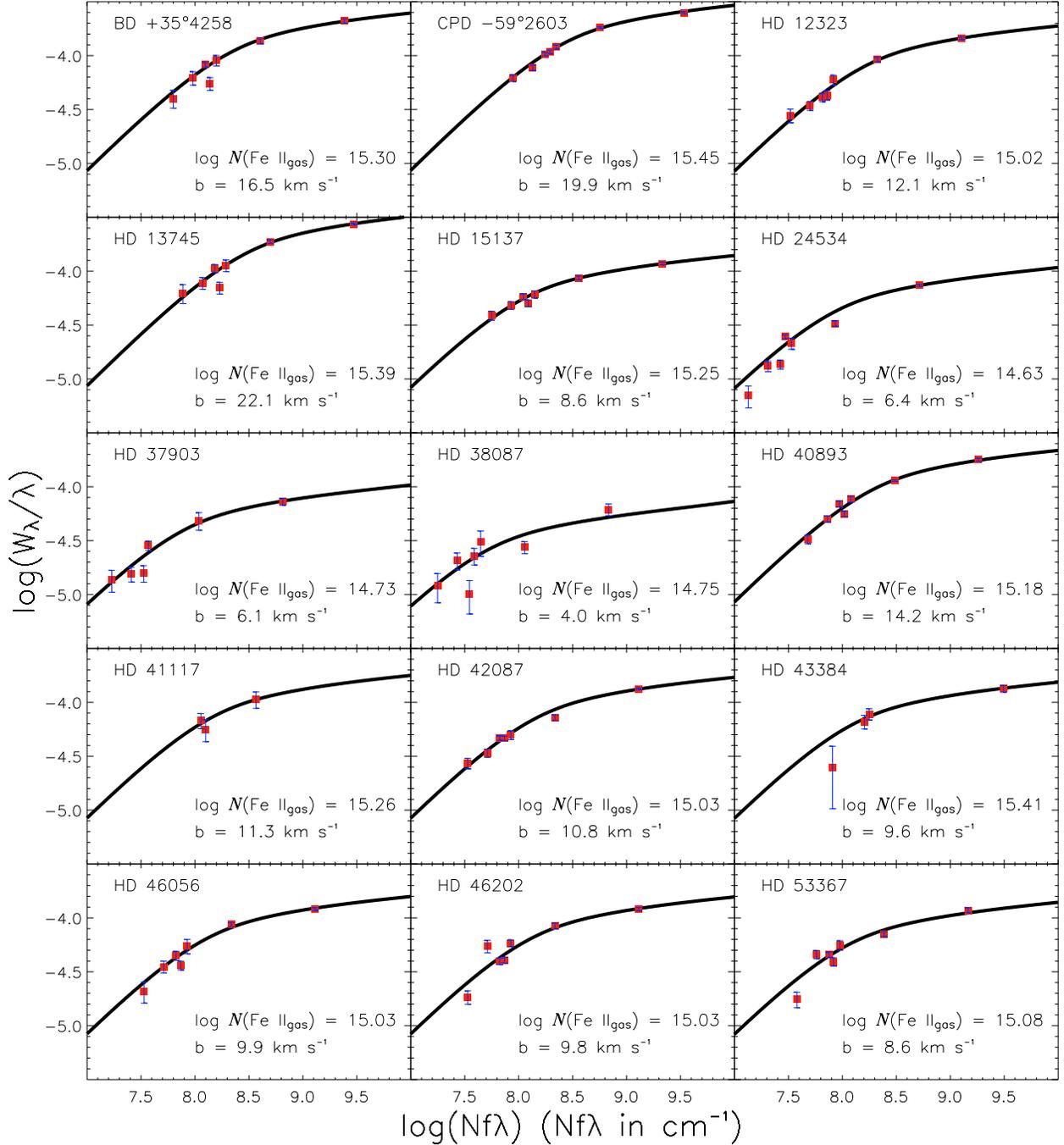}
\end{center}
\caption[Fe II Curves of growth, BD +35$^{\circ}$4258 through HD 53367]{Curves of growth for BD +35$^{\circ}$4258 through HD 53367.  Measured equivalent widths of Fe II absorption lines are plotted on top of appropriate curves of growth for the adopted solution of column density and $b$-value.  Equivalent width measurements are red squares with blue error bars.  The choice of damping constant used to construct the curves has only a trivial effect.}
\label{fig:cogs1-15}
\end{figure}

\begin{figure}
\begin{center}
\epsscale{1.00}
\plotone{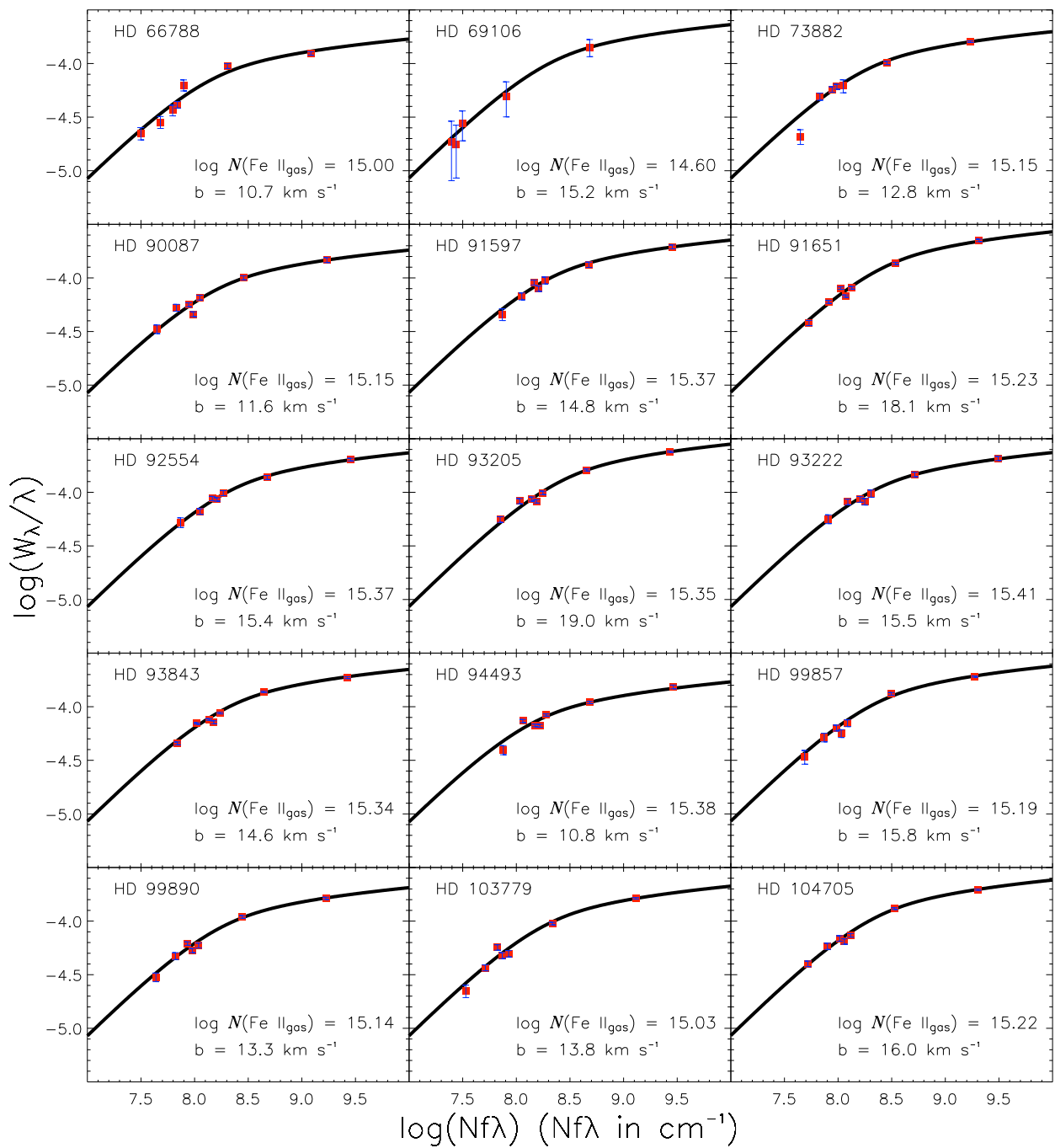}
\end{center}
\caption[Fe II Curves of growth, HD 66788 through HD 104705]{Curves of growth for HD 66788 through HD 104705.  Symbols and other details are the same as in Figure \ref{fig:cogs1-15}.}
\label{fig:cogs16-30}
\end{figure}

\begin{figure}
\begin{center}
\epsscale{1.00}
\plotone{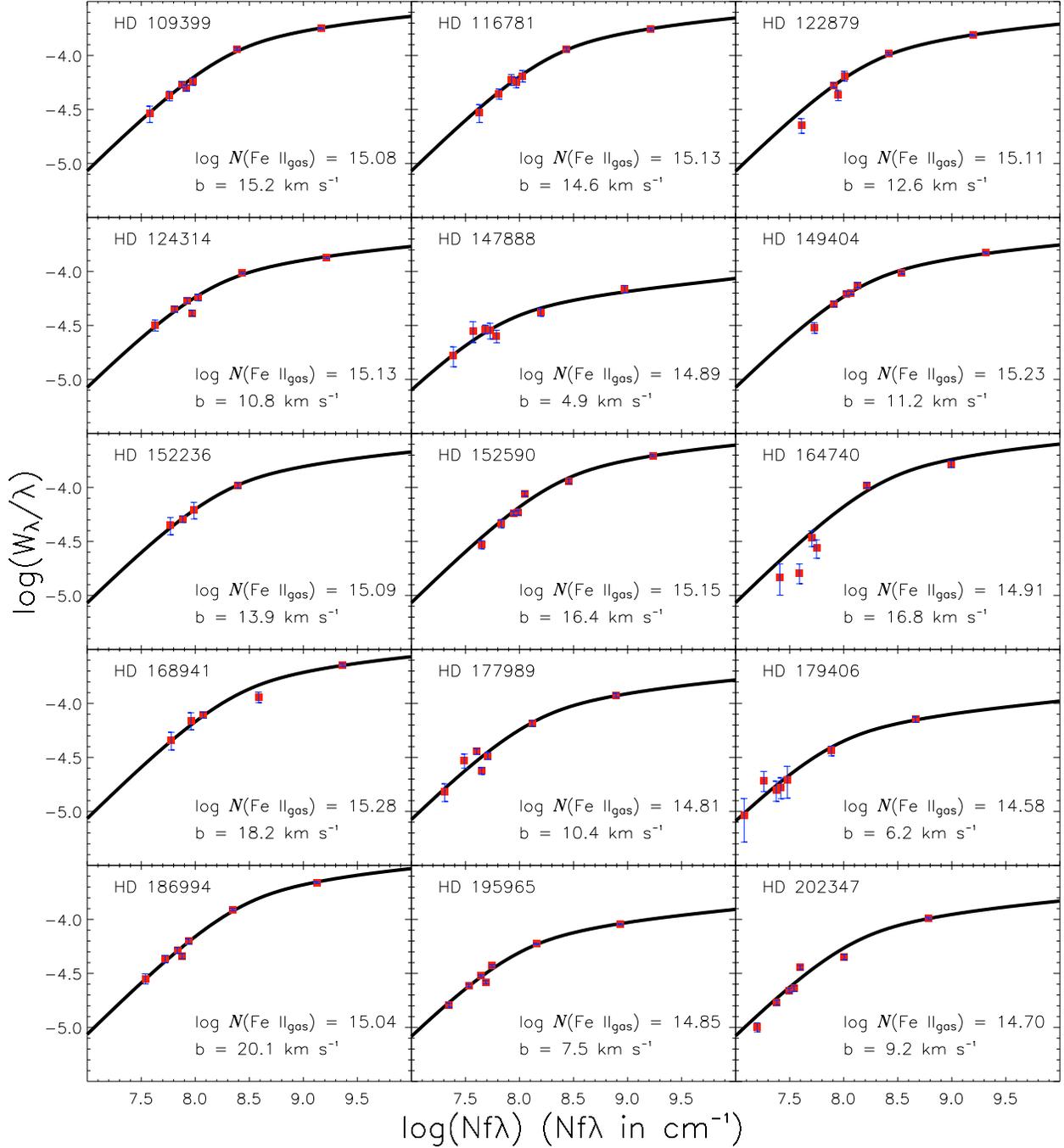}
\end{center}
\caption[Fe II Curves of growth, HD 109399 through HD 202347]{Curves of growth for HD 109399 through HD 202347.  Symbols and other details are the same as in Figure \ref{fig:cogs1-15}.}
\label{fig:cogs31-45}
\end{figure}

\begin{figure}
\begin{center}
\epsscale{1.00}
\plotone{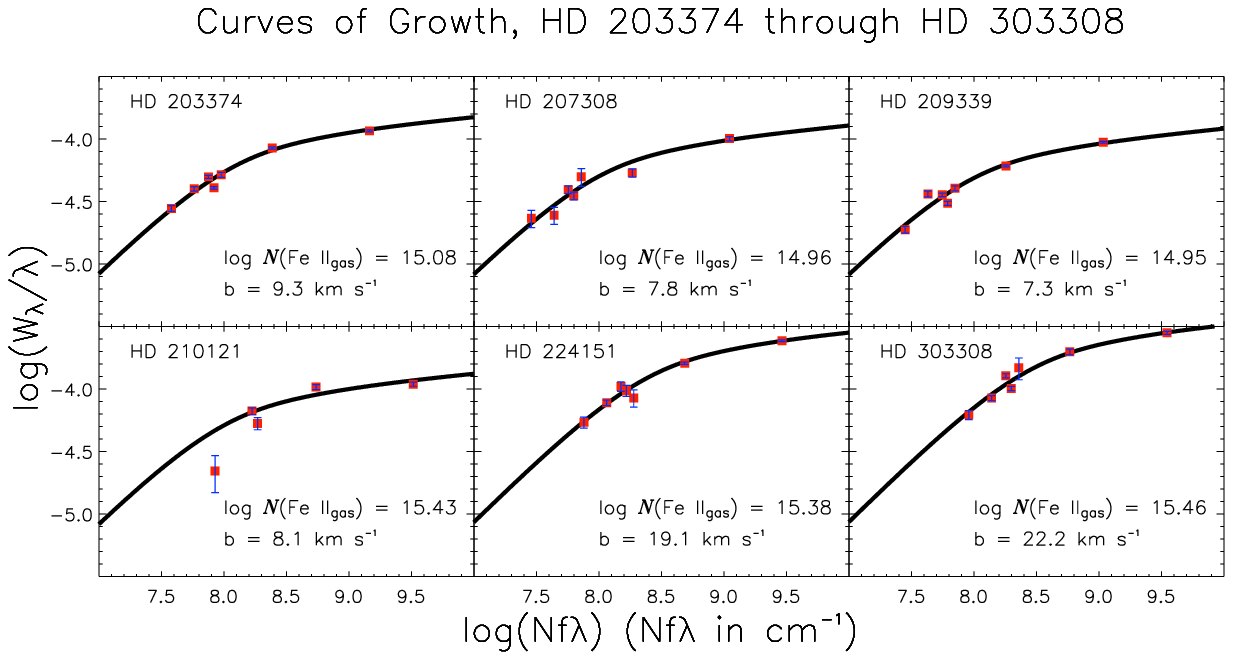}
\end{center}
\caption[Fe II Curves of growth, HD 203374 through HD 303308]{Curves of growth for HD 203374 through HD 303308.  Symbols and other details are the same as in Figure \ref{fig:cogs1-15}.}
\label{fig:cogs46-51}
\end{figure}

\begin{figure}
\begin{center}
\epsscale{1.00}
\plotone{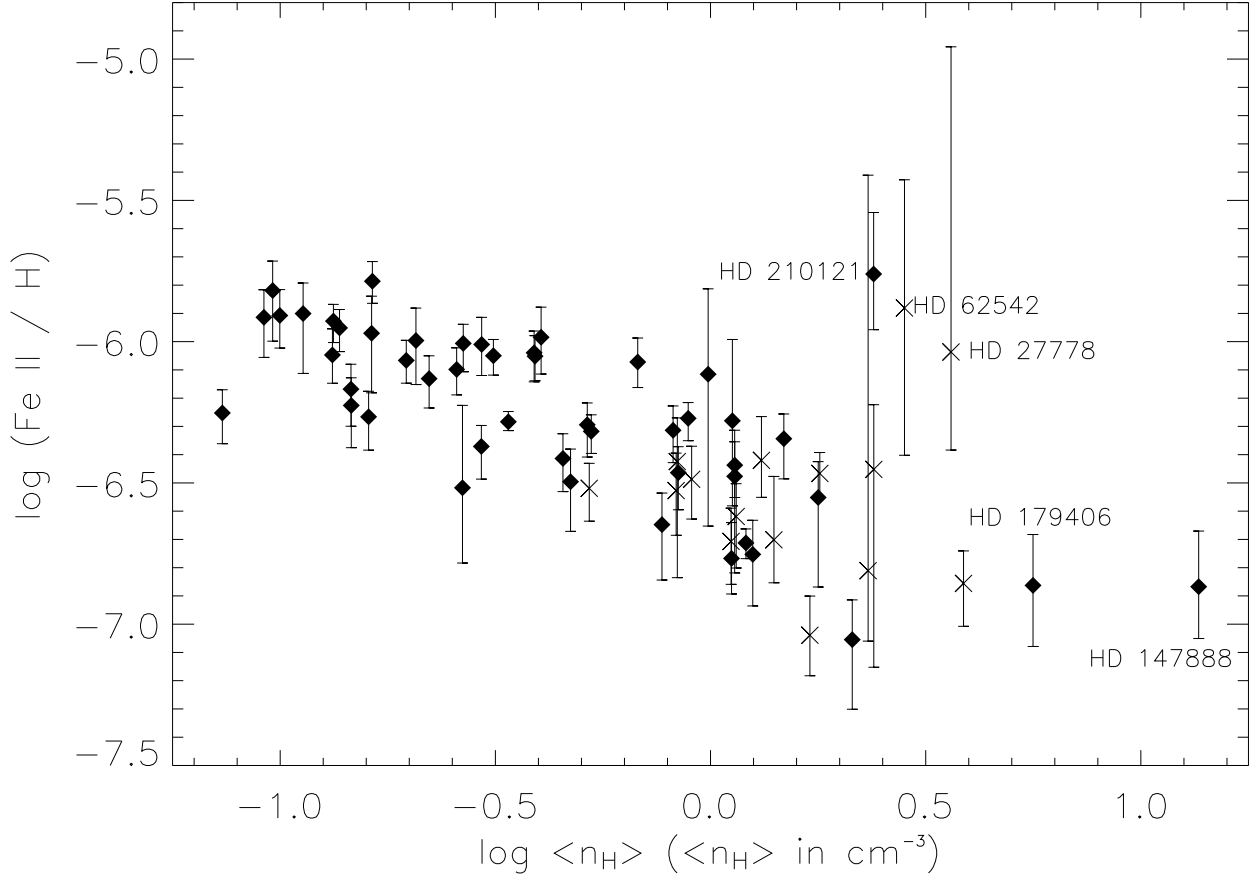}
\end{center}
\caption[$\logFeIIH$ vs. $\log{\nHavg}$]{The logarithmic abundance of Fe II relative to hydrogen plotted against the logarithm of the average volume density of hydrogen.  {\bf Solid diamonds}:---abundances derived in this paper.  {\bf X's}:---abundances based on column densities derived in SRF2002.  The three lines of sight that do not follow the general trend of increased depletion with increased average volume density of hydrogen are HD 210121 from this paper and HD 27778 and HD 62542 from SRF2002.  Note that the two densest lines of sight, HD 179406 and HD 147888, follow the trend but do not show significantly enhanced iron depletion.}
\label{fig:logFeIIHlognh}
\end{figure}

\begin{figure}
\begin{center}
\epsscale{1.00}
\plotone{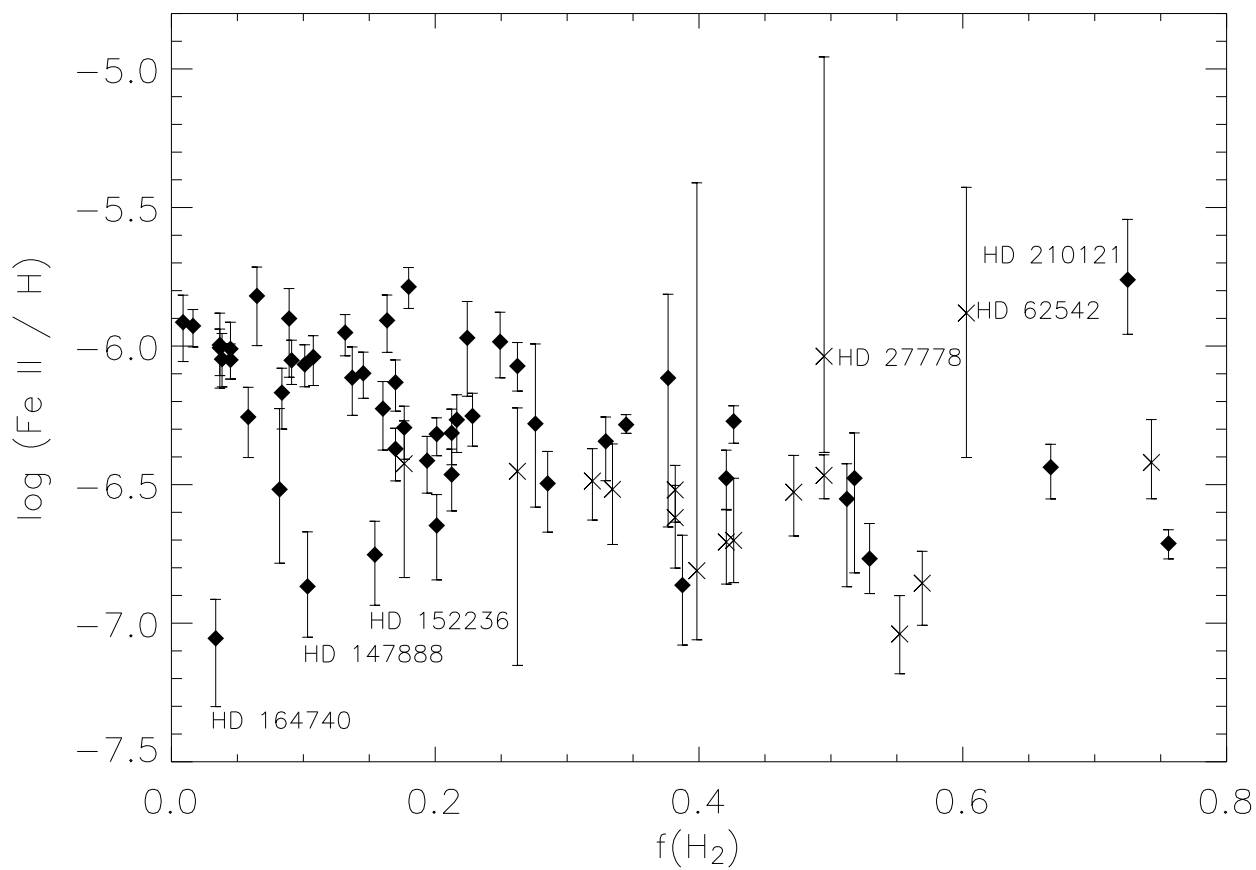}
\end{center}
\caption[$\logFeIIH$ vs. $\fHmol$]{The logarithmic abundance of Fe II relative to hydrogen plotted against the molecular fraction of hydrogen.  Symbols the same as in Figure \ref{fig:logFeIIHlognh}.  While there is some scatter at both ends, there is a strong overall correlation between depletion and $\fHmol$.  The major outlying lines of sight are HD 27778, HD 62542, and HD 210121 with large $\fHmol$ but relatively small depletions compared to the rest of the sample (cf. Figure \ref{fig:logFeIIHlognh}) and HD 147888, HD 152236, and HD 164740 with large depletions but small $\fHmol$.}
\label{fig:logFeIIHHf}
\end{figure}

\begin{figure}
\begin{center}
\epsscale{1.00}
\plotone{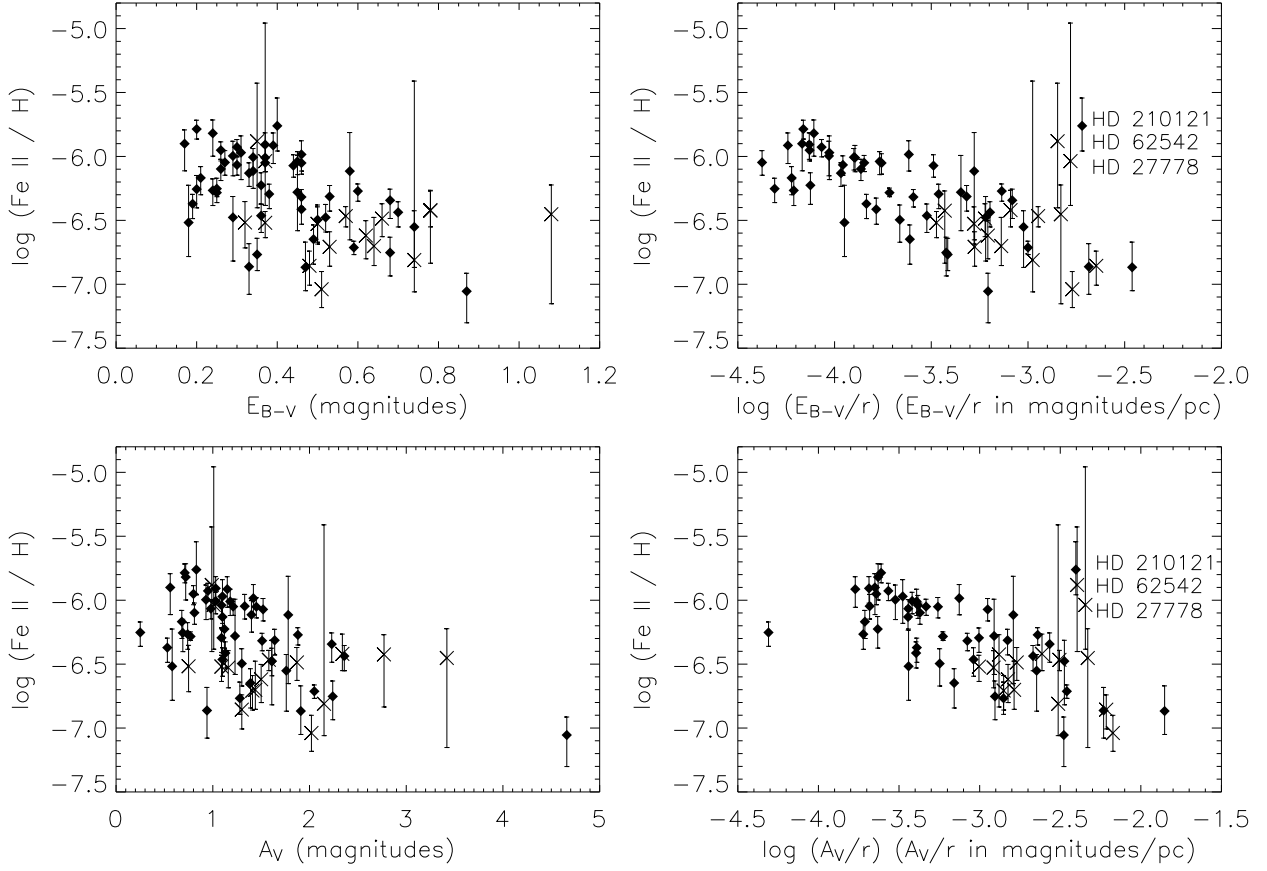}
\end{center}
\caption[$\logFeIIH$ vs. $\log{\ebvdist}$]{The logarithmic abundance of Fe II relative to hydrogen plotted against several measures of reddening and extinction.  {\bf Upper left}:---$\ebv$.  {\bf Upper right}:---$\ebvdist$.  {\bf Lower left}:---$\av$.  {\bf Lower right}:---$\avdist$.  Symbols the same as in Figure \ref{fig:logFeIIHlognh}.  Note the following:  (1) similar to Figures \ref{fig:logFeIIHlognh} and \ref{fig:logFeIIHHf}, the lines of sight toward HD 27778, HD 62542, and HD 210121 have larger values of extinction and reddening per unit pathlength but do have smaller depletions that break from the observed trend and (2) the densest lines of sight, while adhering to the trend, do not show extreme depletions.}
\label{fig:logFeIIHred}
\end{figure}

Figures \ref{fig:cogs1-15}-\ref{fig:cogs46-51} show significant anecdotal evidence that the cases with the smallest errors in the equivalent widths produce extremely self-consistent curves of growth using the \citet{Howk} $f$-values.  The same holds true, in large part, for the curves of growth in SRF2002.  We have carried through our curve-of-growth method using both the \citet{Howk} and \citet{RU1998}.  In 39 of 51 cases, using the \citet{RU1998} $f$-values produces values of $\chi^2$ that are larger than the $\chi^2$ values that result from using the \citet{Howk} values.  When comparing the 102 cases (51 lines of sight using each set of $f$-values), the 10 poorest fits use the \citet{RU1998} $f$-values.  It is also worth noting that in several cases the \citet{RU1998} $f$-values produce a column density of Fe II that is unreasonably large---more than an order of magnitude larger than any of the column densities that are produced using the \citet{Howk} $f$-values, and producing gas-phase abundances on the order of $10^{-5}$, much larger than has been historically observed.  Though some of the initial calculations using the \citet{Howk} $f$-values similarly produce very large column densities, an alternate solution (i.e., values in the $\chi^2$ array that are separated in parameter space from the best solution, but where $\chi^2$ is below the 1-$\sigma$ cutoff) always presents itself.  This is not true of the calculations using the \citet{RU1998} $f$-values, where 15 of 51 lines of sight (including 4 of the 12 cases where the \citeauthor{RU1998} $f$-values improve $\chi^2$ compared to using the \citeauthor{Howk} $f$-values) fail to have a reasonable solution.

In another four of the 12 cases where use of the \citet{RU1998} $f$-values improves the $\chi^2$ compared to the \citet{Howk} $f$-values the improvement is a factor of two or more.  In these cases, however, the reduced $\chi^2$ values using the \citet{Howk} $f$-values are $\lesssim1$, with a maximum of 1.37 for HD 147888.  The other cases (HD 41117, HD 168941, and HD 179406) generally have much larger errors in the equivalent widths (contributing to the small values of $\chi^2$) and less than the full complement of lines measured (e.g.,~only three lines are measured for HD 41117).

Figure \ref{fig:cogsfval} shows a sample comparison of the best-fit curves of growth for HD 195965 using the \citet{Howk} $f$-values and the \citet{RU1998} $f$-values.  We have selected HD 195965 because the error in the column density using the \citet{Howk} $f$-values is among the smallest in our sample ($^{+0.03}_{-0.02}\dex$); in addition, this line of sight shows very little evidence of a multiple-component velocity structure.  Examining Figures \ref{fig:cogs1-15}-\ref{fig:cogs46-51}, shows that there are many, many cases where the \citet{Howk} $f$-values produce self-consistent curves of growth, particularly when the equivalent width errors are small.  It is very unlikely that the curves of growth would be this consistent across a wide range of lines of sight if the \citet{Howk} $f$-values were not roughly correct.

It is beyond the scope of this paper to present a more formal analysis of the $f$-values than what has been presented above (though we further discuss one specific exception below).  We should note that the \citet{Howk} and \citet{RU1998} $f$-values are consistent with each other within their errors.  However, the difference is nontrivial for constructing a curve of growth, as can be seen in Figure \ref{fig:cogsfval}.  The discrepancy in the two sets of $f$-values for the seven absorption lines in this study ranges from $\approx5\%$ to a factor of 2.5; on average, the difference is 30\%, or $0.11\dex$.  The consistency of our best curves of growth, however, argues that the \citet{Howk} $f$-values are much more accurate than this, perhaps to within 0.02 or $0.03\dex$ (5-7\%).  However, this is more of an argument for the self-consistency of the \citet{Howk} $f$-values, rather than a strict constraint on the accuracy of their collective magnitude.  A uniform error in the magnitude of the $f$-values would affect our derived column densities but would not significantly affect the relative correlations discussed in \S \ref{ss:FeII_correlations}.

The main exception to our argument for the self-consistency of the \citet{Howk} $f$-values is the 1112 \AA{} line.  Using these $f$-values implies that the 1112 \AA{} line should have a larger $\eqw$ than the 1133 \AA{} line by 9\% in the linear case (where $\eqw \propto f\lambda^2$).  In the 15 lines of sight where SRF2002 measured both lines, $W_{1133} > W_{1112}$ in 11 cases, though the 1-$\sigma$ errors in $\eqw$ do overlap in 14 of the 15 cases ($W_{1133} > W_{1112}$ for HD 197512 even when errors are taken into account).  However, the weighted averages tell a different story:  $W_{1133}=33\pm1$ m\AA{}, while $W_{1112}=27\pm1$ m\AA{}.

We find essentially the same result in our data.  Figures \ref{fig:cogs1-15}-\ref{fig:cogs46-51} show that in the cases with the smallest errors in the equivalent widths and the most self-consistent curve of growth, the 1112 \AA{} line nearly always deviates from the curve of growth.  Our weighted averages are $W_{1133}=49\pm1$ m\AA{} and $W_{1112}=46\pm1$ m\AA{} (the smaller difference between the two in our sample may be due to the fact that with increased $\eqw$, the 1133 \AA{} line is slightly saturated in more cases).  Interestingly, \citet{RU1998} did find that $f_{1112} < f_{1133}$.  Again, we chose HD 195965 to examine this effect.  If we fit a curve of growth while omitting the 1112 \AA{} line, we find a slightly revised column density ($\logFeII=14.87\pm0.03$ as opposed to $\logFeII=14.85^{+0.03}_{-0.02}$).  If we then fit the 1112 \AA{} line to this curve, we find that $f_{1112}=4.61\times10^{-3}$, in much better agreement with \citeauthor{RU1998} than \citeauthor{Howk}~(see Table \ref{linetable}).

In test cases, revising the $f$-value of the 1112 \AA{} line does not significantly alter our curve-of-growth results; therefore, we have not reanalyzed our column densities in light of this potential $f$-value revision.

We are also presenting the first published detections in interstellar absorption of the 1901.773 \AA{} line of Fe II.  Therefore, we can use our results to constrain the $f$-value of this line.  If $\eqw$ and $\lambda$ are in \AA{} and $N$ is in cm$^{-2}$, then the following equation applies:
\begin{equation}\label{eq:weakeqw}
\eqw=8.85\times10^{-21}Nf\lambda^2
\end{equation}
Solving for the $f$-value and substituting in the wavelength of the 1901.773 \AA{} line, we find that $f=\eqw/[3.20\times10^{-14}N]$.  Considering both the errors in $\eqw$ and $\logFeII$, our derived $f$-value is $6.9\pm1.6\times10^{-5}$ for HD 24534 and $5.5\pm2.4\times10^{-5}$ for HD 93222.  The weighted average of these results is $6.5\pm1.3\times10^{-5}$.  This agrees, within the errors, with the \citet{RU1998} theoretical value of $7.00\times10^{-5}$.  Note, however, the assumptions that went into both equivalent width measurements (the choice between fits for the line in the HD 24534 spectrum, and the distinct cloud components assumed for HD 93222).

At the suggestion of the anonymous referee, we have also taken a different approach to calculate the $f$-value of the 1901.773 \AA{} line.  In this approach, we attempt to measure $\eqw$ for the lines even when no line is visibly apparent and there is no feature that will be fit by a program utilizing $\chi^2$ minimization.  This is done by using other lines as a proxy---the 1239.9523 \AA{} line of Mg II \citep[from our work in][]{JensenMgII} in most cases, and the 1608.4511 \AA{} line of Fe II for 24534.  The fits to these lines are examined for the range over which the profile model is less than 99\% of the continuum.  Given the saturation level in these lines, this cutoff value should include nearly all of the equivalent width.  This wavelength range is then translated to the same velocity range near the 1901.773 \AA{} line, and a measurement of $\eqw$ is made by summing the depth of the data points relative to the assumed continuum.  Fluctuations from the noise are assumed to cancel out to zero, and residual equivalent widths can be potentially observed.

To calculate an error on $\eqw$, we adapt the formula presented by \citet{Jenkins1973} for the maximum equivalent width of an undetected absorption line:
\begin{equation}\label{eq:upperlimits}
W_{\lambda,max}=\frac{N_{\sigma} d\lambda \sqrt{M}}{\rm S/N}
\end{equation}
In this equation, $W_{\lambda,max}$ is the upper limit on $\eqw$, $N_{\sigma}$ is the number of $\sigma$ confidence desired, $d\lambda$ is the wavelength spacing of the pixels, $M$ is the number of consecutive pixels required for detection, and S/N is the signal-to-noise of the local continuum.  We use translate this upper limit on the equivalent width into an error on the equivalent widths that we measure.  In other words, we assume $W_{\lambda,max}=\sigma(W_{\lambda})$ where the $W_{\lambda,max}$ term is evaluated with $N_{\sigma}=1$.

Even though the resulting values of $\eqw$ do not correspond to visually apparent features and are not statistically significant in many cases, the goal of this procedure is to use these measurements to derive an $f$-value for the line.  The $f$-value is calculated in each case using Equation \ref{eq:weakeqw}.  The error in the $f$-value is carried through using standard error propagation, considering errors in both $\eqw$ and column density.  The number of pixels used in determining the error from Equation \ref{eq:upperlimits} is the number of pixels that meet the condition mentioned above, that the range where the model fit is less than 99\% of the continuum.

Within this alternate method, we approach the problem in two different ways.  The first way is to perform the summation over the entire range where all components of the proxy fit are less than 99\%, calculating the $f$-value based on the column density derived through the {\it FUSE} lines.  In this case, the weighted average is $3.1^{+1.2}_{-1.1}\times10^{-5}$.  The other way is to perform the summation over only the velocity range where the dominant component is less than 99\% of the continuum.  We then derive the $f$-value using an adjusted column density, using the Fe II column densities of this paper (or SRF2002), but scaled by the relative fraction of the column density found in the dominant component of the proxy fit.  Using this method, the weighted average of the fits is $6.5\pm1.5\times10^{-5}$.  The fact that summing over the larger range results in a smaller weighted average shows that either (1) the noise fluctuations over such small ranges do not cancel on average or (2) there is some error in fitting the continuum.

We prefer the method of directly analyzing the lines that are actually observed and fit with statistical significance.  However, the alternate method of summing over the data points is also in rough agreement with the previously calculated value of the lines $f$-value, particularly when we only analyze the dominant component.  Nevertheless, this method depends on calculating a statistically significant average from statistically insignificant individual measurements, and is subject to systematic error of making an assumption about what range over which the summation should be performed.  In either case, as stated in \S \ref{ss:1902line}, we have made an important step in roughly confirming the $f$-value of this line.  However, further study to improve these constraints is important so that this line can be used in future studies with COS.

\subsection{The Cosmic Value of Fe/H}
\label{ss:cosmic_iron}
In the past several years, many papers have attempted to summarize the current knowledge of stellar and solar abundances, with implications for the cosmic abundance ``standards'' in the ISM, if consistent standards in fact exist.  Three papers of particular importance for this discussion are \citet{SnowWitt}, \citet{SofiaMeyer}, and \citet{Lodders}.

\citet{SnowWitt} argued for a cosmic iron abundance (including all iron in both gas and dust) of $\logFeH=-4.57$, based on an average of field B stars ($\logFeH=-4.51$), cluster B stars ($\logFeH=-4.49$), and disk F and G stars ($\logFeH=-4.74$).  For iron (along with many other elements), \citeauthor{SnowWitt} concluded that the Sun was substantially overabundant ($\logFeH=-4.33$), an anomalous data point relative to the true Galactic abundances.

\citeauthor{SofiaMeyer}, examined B stars (without differentiation between cluster or field stars) and disk F and G stars, and found a weighted average of $\logFeH=-4.55$ for each sample.  The B star abundances of both \citeauthor{SnowWitt} and \citeauthor{SofiaMeyer} are reasonably consistent, but with a significant difference between the F and G star abundances; \citeauthor{SofiaMeyer} explain this difference as a result of restricting their sample to stars with ages $\leq 2$ Gyr.  In any case, the final adopted cosmic iron abundance of both papers is consistent to with 0.02$\dex$, well within the errors.  Contrary to \citeauthor{SnowWitt}, however, \citet{SofiaMeyer} argued that the solar abundances may in fact reasonably represent cosmic abundances for many elements, in part due to downward revisions to solar abundances \citep[mainly][]{Holweger}.


\citet{Lodders} summarized CI chondritic abundances and solar photospheric abundances, and used both to derive protosolar abundances.  \citeauthor{Lodders} adopts a protosolar iron abundance of $\logFeH=-4.46$.  The overall agreement (0.10$\dex$---25\%---not including the errors) between all types of measurements (stellar and solar) is the among the best for any of the elements discussed in \citeauthor{SnowWitt} and \citeauthor{SofiaMeyer}.  Given this agreement, we can infer the amount of iron in dust fairly confidently.  We adopt a cosmic abundance of iron of $3.1\pm^{+2}_{-1}\times10^{-5}$ (a weighted average of both \citeauthor{SofiaMeyer} measurements and the \citeauthor{Holweger} solar abundance).   Our weighted average implies that only $\sim$1\% of the cosmic abundance in the gas-phase, and our maximum gas-phase abundances are at most approximately $2\times10^{-6}$.  This places a very tight constraint on iron that can be used in dust models \citep*[e.g.][]{ZDA2004}.  However, there are reasons to question any cosmic abundance standard, as we have discussed previously \citep{Snow2000, JensenOI, JensenNI}---many processes, particularly those in stellar formation, could cause abundances in stars to deviate from the abundances of the ISM.

SB1979 found some significant spatial variations in the Fe/H abundance---namely that Scorpius-Ophiuchus lines of sight were substantially more depleted than two lines of sight in Cygnus.  In our sample, we do not note any particularly strong spatial effects.  Most of our lines of sight are in or very near the disk ($|b|<10^{\circ}$), and within that range there is no hint of a correlation between iron depletion and Galactic $b$.  There are slight trends with Galactic longitude $l$ and overall location; in particular, many of the stars in the Crux/Musca region are somewhat less depleted ($\logFeH \approx -6$) than the average of our sample, though this is not true of the more reddened (and nearby) HD 110432 in the SRF2002 sample.  However, this deviation is not substantial compared to the overall scatter in our sample.


\section{SUMMARY}
\label{s:FeII_summary}
Iron abundances and depletions were explored in the past by SB1979 and JSS1986 with {\it Copernicus} and more recently by SRF2002 with {\it FUSE}.  We have undertaken a survey that covers a wider range of reddening and extinction than covered in any of these studies.  We find evidence that iron depletion correlates with many line of sight parameters (such as $\nHavg$, $\ebv$, $\ebvdist$, $\av$, $\avdist$, and $\fHmol$).  Some of these correlations have been noted previously while others have not.  The fact that we observe trends that have not been previously observed may be explained by the following:  (1) this is the largest survey of interstellar Fe II yet performed with {\it FUSE} and, in conjunction with data points from SRF2002, probes the widest range in line of sight parameters in a self-consistent manner; (2) {\it Copernicus} suffered from scattered light to a greater degree than {\it FUSE} (of order 10\% for {\it Copernicus} compared to only a few percent for {\it FUSE}), which may be responsible, to a limited degree, for some scatter in the measured iron abundances; (3) many {\it Copernicus} lines of sight had very short pathlengths, over which intrinsic scatter may be significant; and (4) quite simply, the {\it Copernicus} lines of sight have not been analyzed in the light of some of these parameters, especially $\av$.  Correlations between iron depletion and $\ebv$, $\ebvdist$, $\av$, $\avdist$, and $\fHmol$ can all be interpreted as related to the correlation between iron depletion and the average line of sight density $\nHavg$.  While a few of our lines of sight probe slightly larger density and/or extinction than previous studies, we do not see depletions that are particularly extreme relative to previously observed depletions, suggesting that at best our lines of sight only partially probe true translucent clouds, and are instead dominated by lines of sight with integrated ``translucent'' levels of reddening and extinction.

Also of note are two detections of the 1901.773 \AA{} line of Fe II in the spectra of HD 24534 and HD 93222.  This very weak line is potentially very important for determining iron abundances in a straightforward fashion, without fear of saturation, with the Cosmic Origins Spectrograph.  Further detections are needed to better constrain the $f$-value, but our detections make clear that the previously calculated theoretical value \citep{RU1998} is at least roughly correct, and the line should be detectable at typical Fe II column densities, even in highly-reddened lines of sight with potentially extreme depletions, with the high S/N expected for COS.

Finally, we briefly discuss the $f$-values of the FUV absorption lines of Fe II used in this study.  Though the theoretical $f$-values of \citet{RU1998} and the empirical $f$-values of \citet{Howk} are consistent within the errors, the differences are nontrivial for constructing a curve of growth.  We find that the \citet{Howk} empirical $f$-values produce self-consistent curves of growth in far more cases than the \citet{RU1998} $f$-values.  We find an exception in the case of the 1112 \AA{} line--to obtain self-consistent curves of growth; an $f$-value closer to the \citet{RU1998} $f$-value is preferred.

\acknowledgements

We would like to thank B. L. Rachford, S. V. Penton, B. A. Keeney, and C. W. Danforth for past private communication (regarding line-of-sight measurements, STIS PSF's, and IDL programs), some of which we have cited in our past papers, has been invaluable to us in studying abundances and depletions in the ISM.  We also thank J. M. Shull et al. for the use of their unpublished measurements of molecular hydrogen in many of these lines of sight.  Finally, we thank the anonymous referee for many helpful comments and suggestions that greatly improved the manuscript.

\bibliographystyle{apj}
\bibliography{refs}

\clearpage \clearpage

\begin{deluxetable}{cccccc}
\tablecolumns{6}
\tablewidth{0pc}
\tabletypesize{\tiny}
\tablecaption{Lines of Sight for Fe II: Stellar Data\label{stellardata}}
\tablehead{\colhead{Star Name} & \colhead{Spectral Class} & \colhead{l} & \colhead{b} & \colhead{Distance (pc)} & \colhead{Ref.}}
\startdata
BD +35$^{\circ}$4258 & B0.5Vn & 77.19 & -4.74 & 3100 & 1 \\
CPD -59$^{\circ}$2603 & O7V... & 287.59 & -0.69 & 2630 & 2 \\
HD 12323 & O9V & 132.91 & -5.87 & 3900 & 1 \\
HD 13745 & O9.7II((N)) & 134.58 & -4.96 & 1900 & 1 \\
HD 15137 & O9.5V & 137.46 & +7.58 & 3300 & 2 \\
HD 24534 & O9.5pe & 163.08 & -17.14 & 590 & 3 \\
HD 37903 & B1.5V & 206.85 & -16.54 & 910 & 3 \\
HD 38087 & B5V & 207.07 & -16.26 & 480 & 3 \\
HD 40893 & B0IV & 180.09 & +4.34 & 2800 & 3 \\
HD 41117 & B2Iavar & 189.69 & -0.86 & 1000 & 3 \\
HD 42087 & B2.5Ibe & 187.75 & +1.77 & 1200 & 3 \\
HD 43384 & B3Ia & 187.99 & +3.53 & 1100 & 3 \\
HD 46056 & O8V & 206.34 & -2.25 & 2300 & 3 \\
HD 46202 & O9V & 206.31 & -2.00 & 2000 & 3 \\
HD 53367 & BOIV:e & 223.71 & -1.90 & 780 & 3 \\
HD 66788 & O8/O9Ib & 245.43 & +2.05 & \nodata & \nodata \\
HD 69106 & B1/B2II & 254.52 & -1.33 & 1600 & 3 \\
HD 73882 & O8V: & 260.18 & +0.64 & 1100 & 3 \\
HD 90087 & B2/B3III & 285.16 & -2.13 & 2716 & 2 \\
HD 91597 & B7/B8IV/V & 286.86 & -2.37 & 6400 & 3 \\
HD 91651 & O9VP: & 286.55 & -1.72 & 3500 & 1 \\
HD 92554 & O9.5III & 287.60 & -2.02 & 6795 & 2 \\
HD 93205 & O3V & 287.57 & -0.71 & 2600 & 1 \\
HD 93222 & O7III((f)) & 287.74 & -1.02 & 2900 & 1 \\
HD 93843 & O6III(f) & 228.24 & -0.90 & 2700 & 1 \\
HD 94493 & B0.5Ib & 289.01 & -1.18 & 2900 & 3 \\
HD 99857 & B1Ib & 294.78 & -4.94 & 3058 & 2 \\
HD 99890 & B0.5V: & 291.75 & +4.43 & 3070 & 2 \\
HD 103779 & B0.5II & 296.85 & -1.02 & 3500 & 3 \\
HD 104705 & B0.5III & 297.45 & -0.34 & 3500 & 3 \\
HD 109399 & B1Ib & 301.71 & -9.88 & 1900 & 3 \\
HD 116781 & O9/B1(I)E & 307.05 & -0.07 & \nodata & \nodata \\
HD 122879 & B0Ia & 312.26 & +1.79 & 4800 & 3 \\
HD 124314 & O7 & 312.67 & -0.42 & 1100 & 1 \\
HD 147888 & B3/B4V & 353.65 & +17.71 & 136 & 4 \\
HD 149404 & O9Ia & 340.54 & +3.01 & 820 & 3 \\
HD 152236 & B1Iae & 343.03 & +0.87 & 1800 & 5 \\
HD 152590 & O7.5V & 344.84 & +1.83 & 1800 & 5 \\
HD 164740 & O7.5V & 5.97 & -1.17 & 1400 & 5 \\
HD 168941 & B0III/IV & 5.82 & -6.31 & 5000 & 3 \\
HD 177989 & B2II & 17.81 & -11.88 & 5100 & 3 \\
HD 179406 & B3V & 28.23 & -8.31 & 160 & 3 \\
HD 186994 & BOIII & 78.62 & +10.06 & 2500 & 3 \\
HD 195965 & B0V & 85.71 & +5.00 & 1300 & 3 \\
HD 202347 & B1V & 88.22 & -2.08 & 1300 & 1 \\
HD 203374 & B0IVpe & 100.51 & +8.62 & 820 & 2 \\
HD 207308 & B0.5V & 103.11 & +6.82 & \nodata & \nodata \\
HD 209339 & BOIV & 104.58 & +5.87 & 1100 & 1 \\
HD 210121 & B9 & 56.88 & -44.46 & 210 & 4 \\
HD 224151 & B0.5IISBV & 115.44 & -4.64 & 1355 & 2 \\
HD 303308 & O3V & 287.59 & -0.61 & 2630 & 2 \\
\enddata
\tablerefs{Spectral classes and Galactic coordinates compiled from SIMBAD database at http://simbad.u-strasbg.fr/simbad/sim-fid.  References for distances:  (1) \citet{Savage1985}.  (2) \citet{DS1994}.  (3) Spectroscopic distance modulus from the DIB database at http://dib.uiuc.edu; values compiled from the literature or derived by L.~M.~Hobbs.  (4) Hipparcos parallax of 4-$\sigma$ precision or better.  (5) Member of an OB association, cluster, or multiple-star system, DIB database.}
\end{deluxetable}

\clearpage \clearpage

\begin{deluxetable}{ccccccccc}
\tablecolumns{9}
\tablewidth{0pc}
\tabletypesize{\tiny}
\tablecaption{Lines of Sight for Fe II: Hydrogen Data\label{hydrogentable}}
\tablehead{\colhead{Line Of Sight} & \colhead{$\logHI$} & \colhead{Method\tablenotemark{a}} & \colhead{Ref.} & \colhead{$\logHmol$} & \colhead{Ref.} & \colhead{$\logHtot$} & \colhead{$\nHavg$} & \colhead{$\fHmol$}}
\startdata
BD +35$^{\circ}$4258 & $21.28\pm0.10$ & Ly-$\alpha$ & 1 & $19.56\pm0.03$ & 2 & $21.30\pm0.10$ & $0.21$ & $0.04$ \\
CPD -59$^{\circ}$2603 & $21.46\pm0.07$ & Ly-$\alpha$ & 3 & $20.16\pm0.03$ & 2 & $21.50\pm0.06$ & $0.39$ & $0.09$ \\
HD 12323 & $21.18\pm0.09$ & Ly-$\alpha$ & 4 & $20.32\pm0.08$ & 4 & $21.29\pm0.07$ & $0.16$ & $0.22$ \\
HD 13745 & $21.25\pm0.10$ & Ly-$\alpha$ & 3 & $20.47\pm0.05$ & 2 & $21.37^{+0.08}_{-0.07}$ & $0.40$ & $0.25$ \\
HD 15137 & $21.11\pm0.16$ & Ly-$\alpha$ & 3 & $20.27\pm0.03$ & 2 & $21.22^{+0.13}_{-0.12}$ & $0.16$ & $0.22$ \\
HD 24534 & $20.73\pm0.06$ & Ly-$\alpha$ & 3 & $20.92\pm0.04$ & 5 & $21.34\pm0.03$ & $1.21$ & $0.76$ \\
HD 37903 & $21.17\pm0.10$ & Ly-$\alpha$ & 3 & $20.92\pm0.06$ & 6 & $21.50^{+0.06}_{-0.05}$ & $1.12$ & $0.53$ \\
HD 38087 & $20.91\pm0.30$ & Bohlin & 7 & $20.64\pm0.07$ & 6 & $21.23^{+0.17}_{-0.13}$ & $1.14$ & $0.52$ \\
HD 40893 & $21.50\pm0.10$ & Ly-$\alpha$ & 7 & $20.58\pm0.05$ & 6 & $21.59\pm0.08$ & $0.45$ & $0.19$ \\
HD 41117 & $21.40\pm0.15$ & Ly-$\alpha$ & 3 & $20.68\pm0.30$ & 6 & $21.54^{+0.15}_{-0.13}$ & $1.12$ & $0.28$ \\
HD 42087 & $21.39\pm0.11$ & Ly-$\alpha$ & 3 & $20.52\pm0.12$ & 6 & $21.49\pm0.09$ & $0.84$ & $0.21$ \\
HD 43384 & $21.32\pm0.30$ & Bohlin & 1 & $20.80\pm0.15$ & 6 & $21.53^{+0.21}_{-0.17}$ & $0.99$ & $0.38$ \\
HD 46056 & $21.38\pm0.14$ & Ly-$\alpha$ & 3 & $20.68\pm0.06$ & 6 & $21.53^{+0.11}_{-0.10}$ & $0.47$ & $0.29$ \\
HD 46202 & $21.58\pm0.15$ & Ly-$\alpha$ & 3 & $20.68\pm0.07$ & 6 & $21.68\pm0.12$ & $0.77$ & $0.20$ \\
HD 53367 & $21.32\pm0.30$ & Bohlin & 7 & $21.04\pm0.05$ & 6 & $21.63^{+0.17}_{-0.12}$ & $1.78$ & $0.51$ \\
HD 66788 & $21.23\pm0.10$ & Ly-$\alpha$ & 1 & $19.72\pm0.03$ & 2 & $21.26\pm0.09$ & \nodata & $0.06$ \\
HD 69106 & $21.08\pm0.06$ & Ly-$\alpha$ & 3 & $19.73\pm0.05$ & 2 & $21.12^{+0.06}_{-0.05}$ & $0.27$ & $0.08$ \\
HD 73882 & $21.11\pm0.15$ & Ly-$\alpha$ & 8 & $21.11\pm0.08$ & 5 & $21.59^{+0.08}_{-0.07}$ & $1.14$ & $0.67$ \\
HD 90087 & $21.17\pm0.05$ & Ly-$\alpha$ & 9 & $19.92\pm0.02$ & 9 & $21.22^{+0.05}_{-0.04}$ & $0.20$ & $0.10$ \\
HD 91597 & $21.40\pm0.06$ & Ly-$\alpha$ & 3 & $19.70\pm0.05$ & 2 & $21.42\pm0.06$ & $0.13$ & $0.04$ \\
HD 91651 & $21.15\pm0.06$ & Ly-$\alpha$ & 3 & $19.07\pm0.03$ & 2 & $21.16\pm0.06$ & $0.13$ & $0.02$ \\
HD 92554 & $21.28\pm0.10$ & Ly-$\alpha$ & 3 & $18.93\pm0.05$ & 2 & $21.28\pm0.10$ & $0.09$ & $0.01$ \\
HD 93205 & $21.38\pm0.05$ & Ly-$\alpha$ & 10 & $19.75\pm0.03$ & 2 & $21.40\pm0.05$ & $0.31$ & $0.04$ \\
HD 93222 & $21.40\pm0.07$ & Ly-$\alpha$ & 10 & $19.77\pm0.03$ & 2 & $21.42\pm0.07$ & $0.29$ & $0.04$ \\
HD 93843 & $21.33\pm0.08$ & Ly-$\alpha$ & 3 & $19.61\pm0.03$ & 2 & $21.35\pm0.08$ & $0.27$ & $0.04$ \\
HD 94493 & $21.08\pm0.05$ & Ly-$\alpha$ & 10 & $20.12\pm0.05$ & 11 & $21.17\pm0.04$ & $0.16$ & $0.18$ \\
HD 99857 & $21.24\pm0.08$ & Ly-$\alpha$ & 10 & $20.25\pm0.05$ & 2 & $21.32\pm0.07$ & $0.22$ & $0.17$ \\
HD 99890 & $20.93\pm0.13$ & Ly-$\alpha$ & 3 & $19.47\pm0.05$ & 2 & $20.96\pm0.12$ & $0.10$ & $0.06$ \\
HD 103779 & $21.16\pm0.10$ & Ly-$\alpha$ & 3 & $19.82\pm0.05$ & 2 & $21.20\pm0.09$ & $0.15$ & $0.08$ \\
HD 104705 & $21.11\pm0.07$ & Ly-$\alpha$ & 3 & $19.99\pm0.03$ & 2 & $21.17\pm0.06$ & $0.14$ & $0.13$ \\
HD 109399 & $21.11\pm0.06$ & Ly-$\alpha$ & 3 & $20.04\pm0.20$ & 1 & $21.18\pm0.06$ & $0.26$ & $0.15$ \\
HD 116781 & $21.18\pm0.10$ & Ly-$\alpha$ & 1 & $20.08\pm0.05$ & 2 & $21.24\pm0.09$ & \nodata & $0.14$ \\
HD 122879 & $21.26\pm0.12$ & Ly-$\alpha$ & 4 & $20.24\pm0.09$ & 4 & $21.34\pm0.10$ & $0.15$ & $0.16$ \\
HD 124314 & $21.34\pm0.10$ & Ly-$\alpha$ & 3 & $20.47\pm0.03$ & 2 & $21.44\pm0.08$ & $0.82$ & $0.21$ \\
HD 147888 & $21.71\pm0.09$ & Ly-$\alpha$ & 4 & $20.47\pm0.05$ & 6 & $21.76\pm0.08$ & $13.63$ & $0.10$ \\
HD 149404 & $21.40\pm0.14$ & Ly-$\alpha$ & 3 & $20.79\pm0.04$ & 6 & $21.57^{+0.10}_{-0.09}$ & $1.48$ & $0.33$ \\
HD 152236 & $21.77\pm0.13$ & Ly-$\alpha$ & 3 & $20.73\pm0.12$ & 6 & $21.84\pm0.11$ & $1.25$ & $0.15$ \\
HD 152590 & $21.37\pm0.06$ & Ly-$\alpha$ & 4 & $20.47\pm0.07$ & 4 & $21.47\pm0.05$ & $0.53$ & $0.20$ \\
HD 164740 & $21.95\pm0.15$ & Ly-$\alpha$ & 8 & $20.19\pm0.10$ & 6 & $21.96^{+0.15}_{-0.14}$ & $2.13$ & $0.03$ \\
HD 168941 & $21.11\pm0.09$ & Ly-$\alpha$ & 3 & $20.10\pm0.05$ & 2 & $21.19^{+0.08}_{-0.07}$ & $0.10$ & $0.16$ \\
HD 177989 & $20.95\pm0.09$ & Ly-$\alpha$ & 3 & $20.12\pm0.05$ & 2 & $21.06\pm0.07$ & $0.07$ & $0.23$ \\
HD 179406 & $21.23\pm0.15$ & SvS & 12 & $20.73\pm0.07$ & 6 & $21.44^{+0.10}_{-0.09}$ & $5.61$ & $0.39$ \\
HD 186994 & $20.90\pm0.15$ & Ly-$\alpha$ & 13 & $19.59\pm0.04$ & 6 & $20.94^{+0.14}_{-0.13}$ & $0.11$ & $0.09$ \\
HD 195965 & $20.95\pm0.03$ & Ly-$\alpha$ & 14 & $20.37\pm0.03$ & 2 & $21.13\pm0.02$ & $0.34$ & $0.34$ \\
HD 202347 & $20.99\pm0.10$ & Ly-$\alpha$ & 1 & $20.00\pm0.06$ & 11 & $21.07^{+0.09}_{-0.08}$ & $0.29$ & $0.17$ \\
HD 203374 & $21.11\pm0.09$ & Ly-$\alpha$ & 15 & $20.68\pm0.05$ & 2 & $21.35^{+0.06}_{-0.05}$ & $0.89$ & $0.43$ \\
HD 207308 & $21.20\pm0.10$ & Ly-$\alpha$ & 1 & $20.76\pm0.05$ & 2 & $21.44\pm0.06$ & \nodata & $0.42$ \\
HD 209339 & $21.16\pm0.10$ & Ly-$\alpha$ & 1 & $20.19\pm0.03$ & 2 & $21.24\pm0.08$ & $0.52$ & $0.18$ \\
HD 210121 & $20.63\pm0.15$ & 21 cm & 16 & $20.75\pm0.12$ & 5 & $21.19^{+0.10}_{-0.09}$ & $2.39$ & $0.73$ \\
HD 224151 & $21.32\pm0.08$ & Ly-$\alpha$ & 10 & $20.57\pm0.05$ & 2 & $21.45\pm0.06$ & $0.68$ & $0.26$ \\
HD 303308 & $21.45\pm0.08$ & Ly-$\alpha$ & 3 & $20.23\pm0.05$ & 2 & $21.50\pm0.07$ & $0.39$ & $0.11$ \\
\enddata
\tablenotetext{a}{Methods for H I:  Ly-$\alpha$---Profile fitting of the Lyman-$\alpha$ line.  Bohlin---$\NHI=5.8\times10^{21}\ebv-2\NHmol$ from relationship in \citet{Bohlin1978} and further confirmed in \citet{Rachford2002}; errors in $\NHI$ assumed to be $\pm0.30$ dex.  SvS---$\NHI=5.2\times10^{21}\ebv$ from \citet{SvS1985}.  21 cm---Radio observations of the 21 cm line.}
\tablerefs{(1) This paper.  (2) J.~M.~Shull et al., in preparation.  (3) \citet{DS1994}.  (4)  \citet{Cartledge2004}.  (5) \citet{Rachford2002}.  (6) B.~L.~Rachford, in preparation.  (7) \citet{JensenNI}.  (8) \citet{FM1990}.  (9) \citet{Hebrard2005}. (10) \citet{Andre2003}.  (11) \citet{SDJ2007}.  (12) \citet{Hanson1992}.  (13) \citet{Bohlin1978}.  (14) \citet{Hoopes2003}.  (15) \citet{DS1994}, Table 2 (lines of sight with uncertain stellar parameters).  (16) \citet{WF1992}.}
\end{deluxetable}

\clearpage \clearpage

\begin{deluxetable}{ccccccc}
\tablecolumns{7}
\tablewidth{0pc}
\tabletypesize{\tiny}
\tablecaption{Lines of Sight for Fe II: Reddening Data\label{reddeningtable}}
\tablehead{\colhead{Sightline} & \colhead{$\ebv$} & \colhead{Ref.} & \colhead{$\av$} & \colhead{Ref.} & \colhead{$\rv$} & \colhead{Ref.} \\ \colhead{} & \colhead{(mag)} & \colhead{} & \colhead{(mag)} & \colhead{} & \colhead{} & \colhead{}}
\startdata
BD +35$^{\circ}$4258 & 0.29 & 1 & 0.93 & 2 & 3.21 & 3 \\
CPD -59$^{\circ}$2603 & 0.46 & 4 & 1.45 & 3 & 3.15 & 5 \\
HD 12323 & 0.24 & 1 & 0.74 & 2 & 3.08 & 3 \\
HD 13745 & 0.46 & 4 & 1.42 & 2 & 3.09 & 3 \\
HD 15137 & 0.31 & 4 & 1.10 & 2 & 3.55 & 3 \\
HD 24534 & 0.59 & 4 & 2.05 & 3 & 3.47 & 6 \\
HD 37903 & 0.35 & 4 & 1.28 & 3 & 3.66 & 6 \\
HD 38087 & 0.29 & 6 & 1.61 & 3 & 5.55 & 6 \\
HD 40893 & 0.46 & 6 & 1.13 & 3 & 2.46 & 6 \\
HD 41117 & 0.45 & 4 & 1.23 & 3 & 2.73 & 6 \\
HD 42087 & 0.36 & 6 & 1.10 & 3 & 3.06 & 6 \\
HD 43384 & 0.58 & 6 & 1.78 & 3 & 3.07 & 6 \\
HD 46056 & 0.50 & 6 & 1.30 & 3 & 2.60 & 6 \\
HD 46202 & 0.49 & 6 & 1.39 & 3 & 2.84 & 6 \\
HD 53367 & 0.74 & 6 & 1.76 & 3 & 2.38 & 6 \\
HD 66788 & 0.20 & 5 & 0.69 & 2 & 3.45 & 3 \\
HD 69106 & 0.18 & 6 & 0.58 & 2 & 3.22 & 3 \\
HD 73882 & 0.70 & 6 & 2.36 & 3 & 3.37 & 6 \\
HD 90087 & 0.30 & 4 & 0.98 & 2 & 3.27 & 3 \\
HD 91597 & 0.27 & 6 & 1.33 & 2 & 4.93 & 3 \\
HD 91651 & 0.30 & 4 & 0.95 & 2 & 3.17 & 3 \\
HD 92554 & 0.39 & 4 & 1.15 & 2 & 2.95 & 3 \\
HD 93205 & 0.37 & 4 & 1.21 & 2 & 3.27 & 3 \\
HD 93222 & 0.37 & 7 & 1.18 & 2 & 3.19 & 3 \\
HD 93843 & 0.34 & 4 & 1.03 & 2 & 3.03 & 3 \\
HD 94493 & 0.20 & 4 & 0.71 & 2 & 3.55 & 3 \\
HD 99857 & 0.33 & 4 & 1.10 & 2 & 3.33 & 3 \\
HD 99890 & 0.24 & 4 & 0.72 & 2 & 3.00 & 3 \\
HD 103779 & 0.21 & 4 & 0.68 & 2 & 3.24 & 3 \\
HD 104705 & 0.26 & 4 & 0.80 & 2 & 3.08 & 3 \\
HD 109399 & 0.26 & 4 & 0.81 & 2 & 3.12 & 3 \\
HD 116781 & 0.34 & 5 & 1.40 & 2 & 4.12 & 3 \\
HD 122879 & 0.36 & 7 & 1.12 & 2 & 3.11 & 3 \\
HD 124314 & 0.53 & 4 & 1.64 & 2 & 3.09 & 3 \\
HD 147888 & 0.47 & 6 & 1.91 & 3 & 4.06 & 6 \\
HD 149404 & 0.68 & 4 & 2.23 & 3 & 3.28 & 6 \\
HD 152236 & 0.68 & 4 & 2.24 & 3 & 3.29 & 6 \\
HD 152590 & 0.46 & 6 & 1.51 & 3 & 3.28 & 6 \\
HD 164740 & 0.87 & 6 & 4.66 & 3 & 5.36 & 6 \\
HD 168941 & 0.37 & 4 & 1.03 & 2 & 2.78 & 3 \\
HD 177989 & 0.25 & 4 & 0.25 & 2 & 1.00 & 3 \\
HD 179406 & 0.33 & 6 & 0.94 & 3 & 2.85 & 6 \\
HD 186994 & 0.17 & 6 & 0.56 & 2 & 3.29 & 3 \\
HD 195965 & 0.25 & 4 & 0.77 & 2 & 3.08 & 3 \\
HD 202347 & 0.19 & 8 & 0.53 & 2 & 2.79 & 3 \\
HD 203374 & 0.60 & 9 & 1.88 & 2 & 3.13 & 3 \\
HD 207308 & 0.52 & 8 & 1.61 & 2 & 3.10 & 3 \\
HD 209339 & 0.38 & 8 & 1.09 & 2 & 2.87 & 3 \\
HD 210121 & 0.40 & 6 & 0.83 & 3 & 2.08 & 6 \\
HD 224151 & 0.44 & 4 & 1.52 & 2 & 3.45 & 3 \\
HD 303308 & 0.45 & 7 & 1.09 & 2 & 2.42 & 3 \\
\enddata
\tablerefs{(1) \citet{Savage1985}.  (2) \citet{Neckel1980}.  (3) Derived from the other two quantities via the relationship $\rv \equiv \av / \ebv$.  (4) \citet{DS1994}.  (5) This paper.  (6) DIB Database at http://dib.uiuc.edu; $\ebv$ compiled by L.~M.~Hobbs, $\rv$ derived by B.~L.~Rachford through polarization or infrared photometry \citep[for method and examples, see][]{Rachford2002}.  (7) \citet{Aiello1988}.  (8) \citet{Garmany1992}.  (9) \citet{DS1994}, Table 2 (lines of sight with uncertain stellar parameters).}
\end{deluxetable}

\clearpage \clearpage

\begin{deluxetable}{cccc}
\tablecolumns{4}
\tablewidth{0pc}
\tablecaption{Fe II Absorption Lines Observed\label{linetable}}
\tablehead{\colhead{Wavelength (\AA{})} & \colhead{\citet{RU1998}} & \colhead{\citet{Howk}} & \colhead{Data Used} \\ \colhead{} & \colhead{Theoretical $f$-value} & \colhead{Empirical $f$-value} & \colhead{}}
\startdata
1055.2617 & $6.15\times10^{-3}$ & $7.5\times10^{-3}$ & {\it FUSE} \\
1112.0480 & $4.46\times10^{-3}$ & $6.2\times10^{-3}$ & {\it FUSE} \\
1127.0984 & $1.12\times10^{-3}$ & $2.8\times10^{-3}$ & {\it FUSE} \\
1133.6654 & $4.72\times10^{-3}$ & $5.5\times10^{-3}$ & {\it FUSE} \\
1142.3656 & $4.01\times10^{-3}$ & $4.2\times10^{-3}$ & {\it FUSE} \\
1143.2260 & $1.92\times10^{-2}$ & $1.77\times10^{-2}$ & {\it FUSE} \\
1144.9379 & $8.30\times10^{-2}$ & $1.06\times10^{-1}$ & {\it FUSE} \\
1901.773 & $7.00\times10^{-5}$ & \nodata & {\it HST} \\
\enddata
\tablecomments{Wavelengths are from \citet{Morton2003} and \citet{RU1998}.  Damping constants are unknown for these lines.}
\end{deluxetable}

\clearpage \clearpage

\begin{deluxetable}{cccccccc}
\tablecolumns{8}
\tablewidth{0pc}
\tabletypesize{\tiny}
\tablecaption{Fe II Measured Equivalent Widths:  {\it FUSE} Data\label{eqwidths}}
\tablehead{\colhead{Line Of Sight} & \colhead{$W_{1055.3}$} & \colhead{$W_{1112.0}$} & \colhead{$W_{1127.1}$} & \colhead{$W_{1133.7}$} & \colhead{$W_{1142.4}$} & \colhead{$W_{1143.2}$} & \colhead{$W_{1144.9}$} \\ \colhead{} & \colhead{(m\AA{})} & \colhead{(m\AA{})} & \colhead{(m\AA{})} & \colhead{(m\AA{})} & \colhead{(m\AA{})} & \colhead{(m\AA{})} & \colhead{(m\AA{})}}
\startdata
BD +35$^{\circ}$4258 & $95.2\pm10.6$ & $61.3\pm8.3$ & $45.1\pm8.6$ & $93.2\pm4.8$ & $71.0\pm10.2$ & $155.6\pm8.7$ & $240.1\pm10.9$ \\
CPD -59$^{\circ}$2603 & $127.5\pm5.5$ & $120.4\pm4.5$ & $69.8\pm5.0$ & $117.2\pm4.3$ & $88.1\pm4.9$ & $208.0\pm3.8$ & $285.5\pm6.9$ \\
HD 12323 & $62.7\pm6.8$ & $47.8\pm4.8$ & $31.4\pm4.6$ & $46.4\pm4.2$ & $39.1\pm3.7$ & $105.3\pm4.7$ & $165.8\pm4.9$ \\
HD 13745 & $119.2\pm14.8$ & $77.9\pm9.8$ & $70.4\pm14.1$ & $120.2\pm10.4$ & $88.5\pm11.0$ & $214.4\pm8.1$ & $313.8\pm6.1$ \\
HD 15137 & $64.3\pm5.0$ & $56.1\pm4.1$ & $43.8\pm4.2$ & $65.9\pm4.1$ & $55.0\pm4.6$ & $99.1\pm3.9$ & $134.7\pm3.9$ \\
HD 24534 & $22.6\pm2.7$ & $27.5\pm1.2$ & $7.9\pm1.8$ & $15.5\pm1.5$ & $15.0\pm1.7$ & $37.4\pm2.1$ & $84.8\pm1.8$ \\
HD 37903 & \nodata & $31.6\pm3.1$ & $15.4\pm3.5$ & $17.8\pm3.1$ & $17.7\pm2.7$ & $55.4\pm10.3$ & $83.1\pm6.1$ \\
HD 38087 & $32.4\pm8.6$ & $25.3\pm4.5$ & $13.5\pm4.1$ & $11.4\pm3.9$ & $23.5\pm4.2$ & $31.4\pm4.0$ & $70.5\pm8.9$ \\
HD 40893 & $81.0\pm5.0$ & $62.4\pm3.7$ & $36.6\pm3.3$ & $78.7\pm4.8$ & $57.7\pm3.9$ & $131.0\pm2.3$ & $204.0\pm6.4$ \\
HD 41117 & \nodata & $62.3\pm14.4$ & \nodata & $76.8\pm12.4$ & \nodata & $121.6\pm21.1$ & \nodata \\
HD 42087 & $52.4\pm4.9$ & $51.5\pm2.5$ & $30.6\pm3.4$ & $52.9\pm2.5$ & $38.7\pm3.4$ & $82.7\pm4.9$ & $152.6\pm2.6$ \\
HD 43384 & \nodata & $86.5\pm10.4$ & $27.9\pm16.2$ & $74.9\pm10.9$ & \nodata & \nodata & $153.1\pm10.9$ \\
HD 46056 & $57.8\pm8.8$ & $40.4\pm3.9$ & $23.4\pm5.1$ & $50.7\pm4.8$ & $40.3\pm5.1$ & $99.2\pm4.3$ & $138.5\pm4.1$ \\
HD 46202 & $61.5\pm4.5$ & $45.3\pm2.8$ & $20.7\pm2.9$ & $44.8\pm3.2$ & $62.4\pm8.3$ & $96.5\pm5.8$ & $138.1\pm5.1$ \\
HD 53367 & $58.8\pm6.0$ & $43.6\pm4.0$ & $19.8\pm3.3$ & $51.6\pm2.8$ & $52.3\pm4.7$ & $81.1\pm4.1$ & $134.7\pm7.1$ \\
HD 66788 & $66.4\pm7.9$ & $45.9\pm3.2$ & $25.1\pm3.3$ & $41.6\pm4.8$ & $32.5\pm4.1$ & $108.5\pm5.7$ & $142.0\pm5.2$ \\
HD 69106 & $29.1\pm9.0$ & $19.5\pm10.0$ & \nodata & $21.0\pm11.9$ & \nodata & $56.8\pm20.4$ & $162.2\pm29.5$ \\
HD 73882 & $65.3\pm9.1$ & $67.6\pm3.6$ & $23.5\pm3.7$ & $65.0\pm3.8$ & $56.3\pm4.3$ & $115.8\pm4.0$ & $182.9\pm2.9$ \\
HD 90087 & $69.2\pm3.5$ & $51.0\pm2.6$ & $37.6\pm3.5$ & $63.7\pm2.6$ & $60.3\pm4.5$ & $114.7\pm5.0$ & $168.2\pm8.9$ \\
HD 91597 & $100.0\pm7.4$ & $88.5\pm5.0$ & $51.6\pm6.5$ & $102.7\pm5.6$ & $77.0\pm5.8$ & $151.8\pm8.7$ & $223.2\pm9.2$ \\
HD 91651 & $85.4\pm2.4$ & $75.3\pm1.8$ & $42.9\pm2.4$ & $91.1\pm2.4$ & $68.4\pm2.2$ & $156.1\pm3.7$ & $256.4\pm6.6$ \\
HD 92554 & $103.8\pm5.6$ & $95.5\pm3.6$ & $59.1\pm6.1$ & $99.2\pm4.7$ & $75.7\pm4.6$ & $157.2\pm6.7$ & $231.3\pm6.6$ \\
HD 93205 & $104.4\pm3.9$ & $91.6\pm3.4$ & $62.9\pm3.8$ & $98.0\pm4.3$ & $94.6\pm5.4$ & $184.7\pm4.2$ & $274.1\pm4.5$ \\
HD 93222 & $102.8\pm8.0$ & $90.9\pm6.3$ & $63.5\pm6.0$ & $97.6\pm6.1$ & $93.2\pm5.7$ & $167.8\pm8.3$ & $237.2\pm10.8$ \\
HD 93843 & $92.1\pm1.6$ & $79.6\pm2.5$ & $51.3\pm1.5$ & $86.2\pm1.6$ & $79.8\pm1.4$ & $156.8\pm2.5$ & $213.6\pm4.2$ \\
HD 94493 & $88.3\pm2.2$ & $73.7\pm2.3$ & $44.5\pm4.3$ & $75.8\pm2.2$ & $84.1\pm2.8$ & $125.3\pm3.0$ & $175.0\pm7.0$ \\
HD 99857 & $74.2\pm5.3$ & $62.6\pm5.3$ & $38.4\pm5.6$ & $71.9\pm4.5$ & $59.2\pm5.5$ & $150.7\pm4.4$ & $218.9\pm3.2$ \\
HD 99890 & $62.6\pm3.3$ & $59.0\pm2.6$ & $33.8\pm3.1$ & $70.0\pm2.6$ & $54.0\pm4.0$ & $125.0\pm3.6$ & $186.7\pm4.6$ \\
HD 103779 & $52.5\pm3.6$ & $53.1\pm3.4$ & $25.1\pm3.3$ & $64.4\pm3.8$ & $41.9\pm2.6$ & $108.5\pm3.0$ & $186.9\pm3.2$ \\
HD 104705 & $77.8\pm3.0$ & $71.8\pm4.1$ & $45.0\pm2.7$ & $78.2\pm4.5$ & $66.6\pm4.5$ & $149.5\pm3.0$ & $224.4\pm3.8$ \\
HD 109399 & $60.7\pm4.7$ & $55.7\pm3.6$ & $32.6\pm5.6$ & $60.9\pm4.3$ & $48.3\pm4.9$ & $131.6\pm3.9$ & $204.1\pm3.2$ \\
HD 116781 & $68.3\pm8.3$ & $63.1\pm7.1$ & $33.2\pm6.3$ & $67.7\pm7.9$ & $50.4\pm5.4$ & $130.7\pm5.2$ & $200.8\pm4.7$ \\
HD 122879 & $68.2\pm7.1$ & $47.7\pm5.3$ & $25.4\pm3.9$ & $60.2\pm2.9$ & \nodata & $118.9\pm3.1$ & $176.3\pm2.5$ \\
HD 124314 & $60.6\pm4.1$ & $45.9\pm2.9$ & $35.9\pm4.2$ & $61.0\pm3.3$ & $51.1\pm3.0$ & $111.0\pm2.4$ & $153.6\pm3.1$ \\
HD 147888 & $26.6\pm3.5$ & $31.6\pm5.3$ & $18.7\pm3.9$ & $33.1\pm3.1$ & $32.2\pm7.0$ & $47.9\pm3.9$ & $78.8\pm6.3$ \\
HD 149404 & $78.9\pm4.3$ & $70.5\pm2.4$ & $34.0\pm3.8$ & $70.8\pm3.3$ & $57.1\pm3.2$ & $111.6\pm3.2$ & $169.9\pm3.7$ \\
HD 152236 & $65.4\pm11.4$ & \nodata & \nodata & $57.7\pm4.2$ & $50.9\pm9.2$ & $119.3\pm6.4$ & \nodata \\
HD 152590 & $92.4\pm4.7$ & $65.1\pm3.2$ & $33.3\pm2.9$ & $65.3\pm2.9$ & $52.9\pm4.6$ & $130.8\pm6.6$ & $225.5\pm4.7$ \\
HD 164740 & \nodata & $30.4\pm5.9$ & $16.7\pm5.4$ & $38.5\pm6.4$ & $18.5\pm3.8$ & $119.9\pm6.4$ & $188.2\pm13.4$ \\
HD 168941 & \nodata & \nodata & $51.4\pm9.6$ & $88.3\pm5.6$ & $79.4\pm14.2$ & $130.7\pm14.6$ & $259.4\pm8.2$ \\
HD 177989 & $34.5\pm2.6$ & $26.4\pm1.9$ & $17.1\pm3.2$ & $40.8\pm2.6$ & $33.8\pm5.1$ & $74.4\pm4.4$ & $136.2\pm6.0$ \\
HD 179406 & $20.8\pm6.8$ & $18.7\pm4.1$ & $10.4\pm4.5$ & $17.8\pm3.8$ & $22.1\pm4.7$ & $41.7\pm4.3$ & $81.3\pm4.7$ \\
HD 186994 & $66.2\pm3.7$ & $50.8\pm2.7$ & $31.9\pm3.4$ & $58.5\pm3.6$ & $49.2\pm3.9$ & $141.2\pm3.9$ & $251.0\pm5.1$ \\
HD 195965 & $39.1\pm1.1$ & $29.1\pm1.0$ & $18.1\pm0.8$ & $33.9\pm0.9$ & $27.9\pm0.9$ & $67.7\pm1.0$ & $103.2\pm1.5$ \\
HD 202347 & $37.9\pm1.6$ & $25.5\pm1.5$ & $11.3\pm1.1$ & $24.9\pm1.2$ & $19.5\pm1.0$ & $51.2\pm2.9$ & $117.3\pm2.9$ \\
HD 203374 & $54.6\pm2.4$ & $45.2\pm1.1$ & $31.4\pm1.6$ & $56.4\pm1.5$ & $45.6\pm1.9$ & $96.8\pm1.8$ & $133.1\pm2.3$ \\
HD 207308 & $52.6\pm8.5$ & $39.0\pm2.7$ & $26.2\pm4.2$ & $44.6\pm3.2$ & $28.1\pm4.4$ & $61.3\pm4.6$ & $115.4\pm4.5$ \\
HD 209339 & $42.5\pm1.9$ & $34.0\pm1.3$ & $21.2\pm1.3$ & $40.5\pm1.3$ & $41.5\pm2.2$ & $69.6\pm1.3$ & $107.9\pm1.8$ \\
HD 210121 & \nodata & $59.0\pm6.5$ & $24.9\pm8.2$ & $75.5\pm4.6$ & \nodata & $118.4\pm5.5$ & $125.1\pm4.6$ \\
HD 224151 & $89.6\pm14.1$ & $107.9\pm10.7$ & $61.0\pm6.3$ & $118.9\pm10.2$ & $88.7\pm4.5$ & $183.8\pm4.2$ & $278.1\pm4.1$ \\
HD 303308 & $156.0\pm30.8$ & $112.2\pm4.3$ & $70.0\pm5.8$ & $144.5\pm5.7$ & $96.7\pm4.3$ & $227.0\pm11.2$ & $322.1\pm9.3$ \\
\enddata
\end{deluxetable}

\clearpage \clearpage

\begin{deluxetable}{cccccccc}
\tablecolumns{8}
\tablewidth{0pc}
\tabletypesize{\tiny}
\tablecaption{Fe II Column Density Results\label{coltable}}
\tablehead{\colhead{Sightline} & \colhead{$\logFeII$} & \colhead{$b$-value} & \colhead{$\logFeIIH$} & \colhead{Fe II/H ($\times 10^7$)} & \colhead{Reduced $\chi^2$} & \colhead{Components\tablenotemark{a}} & \colhead{Comments}}
\startdata
BD +35$^{\circ}$4258 & $15.30^{+0.09}_{-0.08}$ & $16.5^{+2.4}_{-2.2}$ & $-6.00^{+0.12}_{-0.16}$ & $10.1^{+3.1}_{-3.0}$ & $1.94$ & 2 & \nodata \\
CPD -59$^{\circ}$2603 & $15.45^{+0.05}_{-0.04}$ & $19.9\pm1.3$ & $-6.05^{+0.07}_{-0.09}$ & $8.9\pm1.6$ & $1.32$ & 3 & \nodata \\
HD 12323 & $15.02\pm0.07$ & $12.1^{+1.4}_{-1.1}$ & $-6.27^{+0.09}_{-0.12}$ & $5.4\pm1.3$ & $0.73$ & 1 & \nodata \\
HD 13745 & $15.39^{+0.09}_{-0.08}$ & $22.1^{+1.7}_{-1.5}$ & $-5.98^{+0.11}_{-0.13}$ & $10.4^{+2.9}_{-2.7}$ & $2.02$ & 3 & \nodata \\
HD 15137 & $15.25^{+0.10}_{-0.08}$ & $8.6^{+0.9}_{-0.8}$ & $-5.97^{+0.13}_{-0.21}$ & $10.7^{+3.8}_{-4.1}$ & $0.97$ & 1 & \nodata \\
HD 24534 & $14.63\pm0.04$ & $6.4\pm0.4$ & $-6.71^{+0.05}_{-0.06}$ & $1.9\pm0.2$ & $10.02$ & 1 & Alt \\
HD 37903 & $14.73^{+0.12}_{-0.10}$ & $6.1^{+1.5}_{-1.2}$ & $-6.77\pm0.13$ & $1.7^{+0.6}_{-0.4}$ & $1.81$ & 1 & \nodata \\
HD 38087 & $14.75^{+0.14}_{-0.12}$ & $4.0^{+1.1}_{-0.8}$ & $-6.48^{+0.16}_{-0.34}$ & $3.3^{+1.5}_{-1.8}$ & $1.14$ & 1 & Alt \\
HD 40893 & $15.18^{+0.06}_{-0.05}$ & $14.2^{+1.3}_{-1.1}$ & $-6.41^{+0.09}_{-0.12}$ & $3.9\pm0.9$ & $2.53$ & 1 & \nodata \\
HD 41117 & $15.26^{+0.28}_{-0.15}$ & $11.3^{+6.2}_{-3.6}$ & $-6.28^{+0.29}_{-0.30}$ & $5.2^{+4.9}_{-2.6}$ & $0.51$ & 1 & Alt \\
HD 42087 & $15.03^{+0.06}_{-0.05}$ & $10.8\pm0.7$ & $-6.46^{+0.09}_{-0.13}$ & $3.4^{+0.8}_{-0.9}$ & $2.09$ & 1 & \nodata \\
HD 43384 & $15.41^{+0.29}_{-0.16}$ & $9.6\pm1.7$ & $-6.12^{+0.30}_{-0.54}$ & $7.7^{+7.7}_{-5.4}$ & $0.48$ & 1 & Alt \\
HD 46056 & $15.03\pm0.09$ & $9.9^{+1.1}_{-0.9}$ & $-6.50^{+0.12}_{-0.18}$ & $3.2^{+1.0}_{-1.1}$ & $1.71$ & 1 & \nodata \\
HD 46202 & $15.03\pm0.07$ & $9.8^{+1.3}_{-1.1}$ & $-6.65^{+0.11}_{-0.20}$ & $2.3^{+0.7}_{-0.8}$ & $4.43$ & 1 & \nodata \\
HD 53367 & $15.08^{+0.09}_{-0.08}$ & $8.6^{+1.5}_{-1.2}$ & $-6.55^{+0.13}_{-0.32}$ & $2.8^{+1.0}_{-1.5}$ & $3.68$ & 1 & \nodata \\
HD 66788 & $15.00^{+0.08}_{-0.07}$ & $10.7\pm1.3$ & $-6.26^{+0.11}_{-0.15}$ & $5.5\pm1.6$ & $3.18$ & 1 & \nodata \\
HD 69106 & $14.60^{+0.29}_{-0.25}$ & $15.2^{+14.8}_{-6.5}$ & $-6.52^{+0.29}_{-0.27}$ & $3.0^{+2.9}_{-1.4}$ & $0.08$ & 1 & Alt \\
HD 73882 & $15.15\pm0.06$ & $12.8^{+0.8}_{-0.7}$ & $-6.44^{+0.08}_{-0.11}$ & $3.7\pm0.8$ & $1.87$ & 1 & \nodata \\
HD 90087 & $15.15\pm0.06$ & $11.6^{+1.8}_{-1.5}$ & $-6.07^{+0.07}_{-0.08}$ & $8.6\pm1.5$ & $4.80$ & 1 & \nodata \\
HD 91597 & $15.37^{+0.08}_{-0.07}$ & $14.8^{+1.9}_{-1.8}$ & $-6.05^{+0.09}_{-0.10}$ & $9.0^{+2.1}_{-1.8}$ & $1.29$ & 1 & \nodata \\
HD 91651 & $15.23\pm0.03$ & $18.1^{+1.4}_{-1.3}$ & $-5.93^{+0.06}_{-0.08}$ & $11.8^{+1.7}_{-1.9}$ & $9.45$ & 1 & \nodata \\
HD 92554 & $15.37^{+0.06}_{-0.05}$ & $15.4^{+1.4}_{-1.3}$ & $-5.91^{+0.10}_{-0.14}$ & $12.2^{+3.1}_{-3.4}$ & $0.67$ & 1 & \nodata \\
HD 93205 & $15.35\pm0.04$ & $19.0^{+1.1}_{-0.9}$ & $-6.05^{+0.06}_{-0.07}$ & $8.9\pm1.3$ & $3.36$ & 3 & \nodata \\
HD 93222 & $15.41^{+0.08}_{-0.07}$ & $15.5^{+2.2}_{-1.8}$ & $-6.01^{+0.10}_{-0.11}$ & $9.8^{+2.4}_{-2.2}$ & $1.06$ & 2 & \nodata \\
HD 93843 & $15.34^{+0.02}_{-0.03}$ & $14.6^{+0.8}_{-0.7}$ & $-6.01^{+0.07}_{-0.10}$ & $9.9^{+1.7}_{-2.0}$ & $8.11$ & 2 & \nodata \\
HD 94493 & $15.38\pm0.06$ & $10.8^{+1.1}_{-0.9}$ & $-5.79^{+0.07}_{-0.08}$ & $16.4^{+2.8}_{-2.7}$ & $10.15$ & 2 & \nodata \\
HD 99857 & $15.19\pm0.06$ & $15.8^{+0.9}_{-0.8}$ & $-6.13^{+0.08}_{-0.10}$ & $7.4^{+1.5}_{-1.6}$ & $1.86$ & 1 & \nodata \\
HD 99890 & $15.14^{+0.05}_{-0.04}$ & $13.3^{+1.1}_{-1.0}$ & $-5.82^{+0.10}_{-0.18}$ & $15.2^{+4.1}_{-5.1}$ & $3.57$ & 1 & \nodata \\
HD 103779 & $15.03\pm0.05$ & $13.8^{+0.9}_{-0.8}$ & $-6.17^{+0.09}_{-0.13}$ & $6.8^{+1.5}_{-1.8}$ & $3.33$ & 2 & \nodata \\
HD 104705 & $15.22\pm0.04$ & $16.0\pm0.9$ & $-5.95^{+0.07}_{-0.08}$ & $11.2^{+1.8}_{-2.0}$ & $1.34$ & 2 & \nodata \\
HD 109399 & $15.08^{+0.06}_{-0.05}$ & $15.2\pm0.9$ & $-6.10^{+0.08}_{-0.09}$ & $8.0\pm1.5$ & $1.39$ & 2 & \nodata \\
HD 116781 & $15.13^{+0.09}_{-0.07}$ & $14.6^{+1.2}_{-1.3}$ & $-6.11^{+0.11}_{-0.14}$ & $7.7^{+2.2}_{-2.1}$ & $0.23$ & 2 & \nodata \\
HD 122879 & $15.11^{+0.06}_{-0.05}$ & $12.6\pm0.6$ & $-6.23^{+0.10}_{-0.15}$ & $5.9^{+1.5}_{-1.7}$ & $2.46$ & 2 & \nodata \\
HD 124314 & $15.13^{+0.06}_{-0.05}$ & $10.8^{+0.8}_{-0.7}$ & $-6.31^{+0.09}_{-0.11}$ & $4.9\pm1.1$ & $4.94$ & 2 & \nodata \\
HD 147888 & $14.89^{+0.19}_{-0.14}$ & $4.9^{+1.4}_{-1.1}$ & $-6.87^{+0.20}_{-0.18}$ & $1.4^{+0.8}_{-0.5}$ & $0.99$ & 1 & Alt \\
HD 149404 & $15.23\pm0.05$ & $11.2^{+0.9}_{-0.7}$ & $-6.34^{+0.09}_{-0.14}$ & $4.5^{+1.0}_{-1.3}$ & $1.88$ & 1 & \nodata \\
HD 152236 & $15.09^{+0.09}_{-0.08}$ & $13.9^{+4.4}_{-2.8}$ & $-6.75^{+0.12}_{-0.18}$ & $1.8\pm0.6$ & $0.05$ & 1 & Alt \\
HD 152590 & $15.15^{+0.04}_{-0.05}$ & $16.4^{+1.3}_{-1.0}$ & $-6.32^{+0.06}_{-0.08}$ & $4.8^{+0.7}_{-0.8}$ & $4.94$ & 1 & \nodata \\
HD 164740 & $14.91^{+0.10}_{-0.08}$ & $16.8^{+3.9}_{-3.1}$ & $-7.05^{+0.14}_{-0.25}$ & $0.9^{+0.3}_{-0.4}$ & $4.93$ & 1 & \nodata \\
HD 168941 & $15.28^{+0.07}_{-0.06}$ & $18.2^{+1.3}_{-1.4}$ & $-5.91^{+0.09}_{-0.12}$ & $12.4\pm2.9$ & $1.37$ & 1 & Alt \\
HD 177989 & $14.81\pm0.06$ & $10.4^{+1.8}_{-1.3}$ & $-6.25^{+0.08}_{-0.11}$ & $5.6\pm1.2$ & $5.57$ & 1 & \nodata \\
HD 179406 & $14.58^{+0.17}_{-0.15}$ & $6.2^{+1.6}_{-1.2}$ & $-6.86^{+0.18}_{-0.22}$ & $1.4^{+0.7}_{-0.5}$ & $0.54$ & 1 & Alt \\
HD 186994 & $15.04\pm0.04$ & $20.1\pm1.4$ & $-5.90^{+0.11}_{-0.21}$ & $12.6^{+3.6}_{-4.9}$ & $2.88$ & 1 & \nodata \\
HD 195965 & $14.85^{+0.03}_{-0.02}$ & $7.5\pm0.4$ & $-6.28^{+0.04}_{-0.03}$ & $5.2^{+0.5}_{-0.4}$ & $7.19$ & 1 & \nodata \\
HD 202347 & $14.70^{+0.03}_{-0.04}$ & $9.2^{+0.9}_{-0.7}$ & $-6.37^{+0.07}_{-0.12}$ & $4.3^{+0.8}_{-1.0}$ & $10.86$ & 1 & \nodata \\
HD 203374 & $15.08^{+0.03}_{-0.04}$ & $9.3\pm0.5$ & $-6.27^{+0.06}_{-0.08}$ & $5.4^{+0.7}_{-0.9}$ & $12.81$ & 1 & \nodata \\
HD 207308 & $14.96^{+0.09}_{-0.08}$ & $7.8^{+1.1}_{-1.0}$ & $-6.48^{+0.10}_{-0.11}$ & $3.3^{+0.9}_{-0.8}$ & $2.60$ & 1 & Alt \\
HD 209339 & $14.95\pm0.04$ & $7.3^{+0.5}_{-0.4}$ & $-6.29^{+0.08}_{-0.11}$ & $5.1^{+1.0}_{-1.2}$ & $9.25$ & 1 & \nodata \\
HD 210121 & $15.43^{+0.21}_{-0.13}$ & $8.1\pm0.9$ & $-5.76^{+0.22}_{-0.20}$ & $17.4^{+11.3}_{-6.3}$ & $6.75$ & 1 & \nodata \\
HD 224151 & $15.38^{+0.07}_{-0.05}$ & $19.1^{+1.0}_{-1.1}$ & $-6.07\pm0.09$ & $8.5^{+1.8}_{-1.6}$ & $1.11$ & 2 & \nodata \\
HD 303308 & $15.46\pm0.05$ & $22.2^{+2.1}_{-1.8}$ & $-6.04^{+0.08}_{-0.10}$ & $9.1^{+1.8}_{-1.9}$ & $6.15$ & 2 & \nodata \\
\enddata
\tablenotetext{a}{Number of marginally resolved velocity components seen in the line of sight in the {\it FUSE} data.}
\tablenotetext{b}{``Alt'' signifies that curve-of-growth method has two local minima, an artificact of calculating possible solutions over a wide range, but the solution with the best consistency with the observed profiles is selected.}
\end{deluxetable}

\clearpage \clearpage

\begin{deluxetable}{ccccc}
\tablecolumns{5}
\tablewidth{0pc}
\tabletypesize{\scriptsize}
\tablecaption{Measured Equivalent Widths and Upper Limits of the 1901.773 \AA{} of Fe II\label{eqwidths1902}}
\tablehead{\colhead{Line Of Sight} & \multicolumn{2}{c}{Entire Range} & \multicolumn{2}{c}{Dominant Component}\\ \colhead{} & \colhead{$W_{1901.773}$} & \colhead{Error} & \colhead{$W_{1901.773}$} & \colhead{Error}}
\startdata
HD 15137 & 1.16 & 2.84 & 2.32 & 2.52 \\
HD 24534 & \multicolumn{2}{c}{Direct fit} & 0.94$\pm$0.17 \\
         & 0.65 & 0.51 & 1.29 & 0.34 \\
HD 93205 & -0.64 & 2.46 & 2.29 & 1.17 \\
HD 93222 & \multicolumn{2}{c}{Direct fit} & 1.95$\pm$0.76 \\
         & 2.06 & 1.92 & 0.77 & 1.14 \\
HD 93843 & 0.86 & 3.29 & 0.91 & 2.06 \\
HD 94493 & 4.34 & 3.64 & 0.40 & 1.32 \\
HD 99857 & 0.63 & 4.36 & 1.34 & 1.78 \\
HD 103779 & 0.60 & 2.88 & 2.85 & 1.87 \\
HD 109399 & 1.71 & 5.02 & 4.10 & 2.73 \\
HD 116781 & -0.94 & 5.28 & 1.24 & 3.13 \\
HD 124314 & 1.39 & 3.04 & -0.34 & 2.36 \\
HD 195965 & 1.71 & 1.55 & -0.31 & 0.93 \\
HD 203374 & 1.40 & 2.03 & 1.13 & 1.97 \\
HD 206267 & 2.98 & 2.90 & 4.48 & 2.25 \\
HD 209339 & 0.34 & 2.27 & 0.19 & 2.14 \\
HD 210839 & 0.83 & 1.72 & 0.84 & 1.42 \\
HD 303308\tablenotemark{a} & 4.39 & 2.93 & 4.96 & 2.20 \\
\enddata
\tablecomments{As described in \S \ref{ss:f-values}, the equivalent width is directly measured by a summation of the normalized depth of the data points multiplied by the wavelength spacing.  This is done for both the entire velocity range over which a proxy fit is less than 99\% of the continuum level, and for the range over which this is only true of the dominant component.  $f$-values are then derived using Equation \ref{eq:weakeqw}, with errors based on the standard error propagation of the values of $\eqw$ shown here and the column densities from this paper (SRF2002 in the case of HD 206267 and HD 210839).  The more direct fits are shown for HD 24534 and HD 93222.}
\tablenotetext{a}{A potential feature is observed in this line of sight, but the continuum normalization is unclear and an attempt at a direct fit is only marginally significant (just over 1-$\sigma$).  The error listed here is smaller than the error of the unrecorded fit because the S/N is calculated over a slightly wider range.}
\end{deluxetable}

\clearpage \clearpage

\clearpage \clearpage

\clearpage \clearpage

\clearpage \clearpage

\clearpage \clearpage

\clearpage \clearpage

\clearpage \clearpage

\clearpage \clearpage

\clearpage \clearpage

\begin{figure}
\begin{center}
\epsscale{1.00}
\plotone{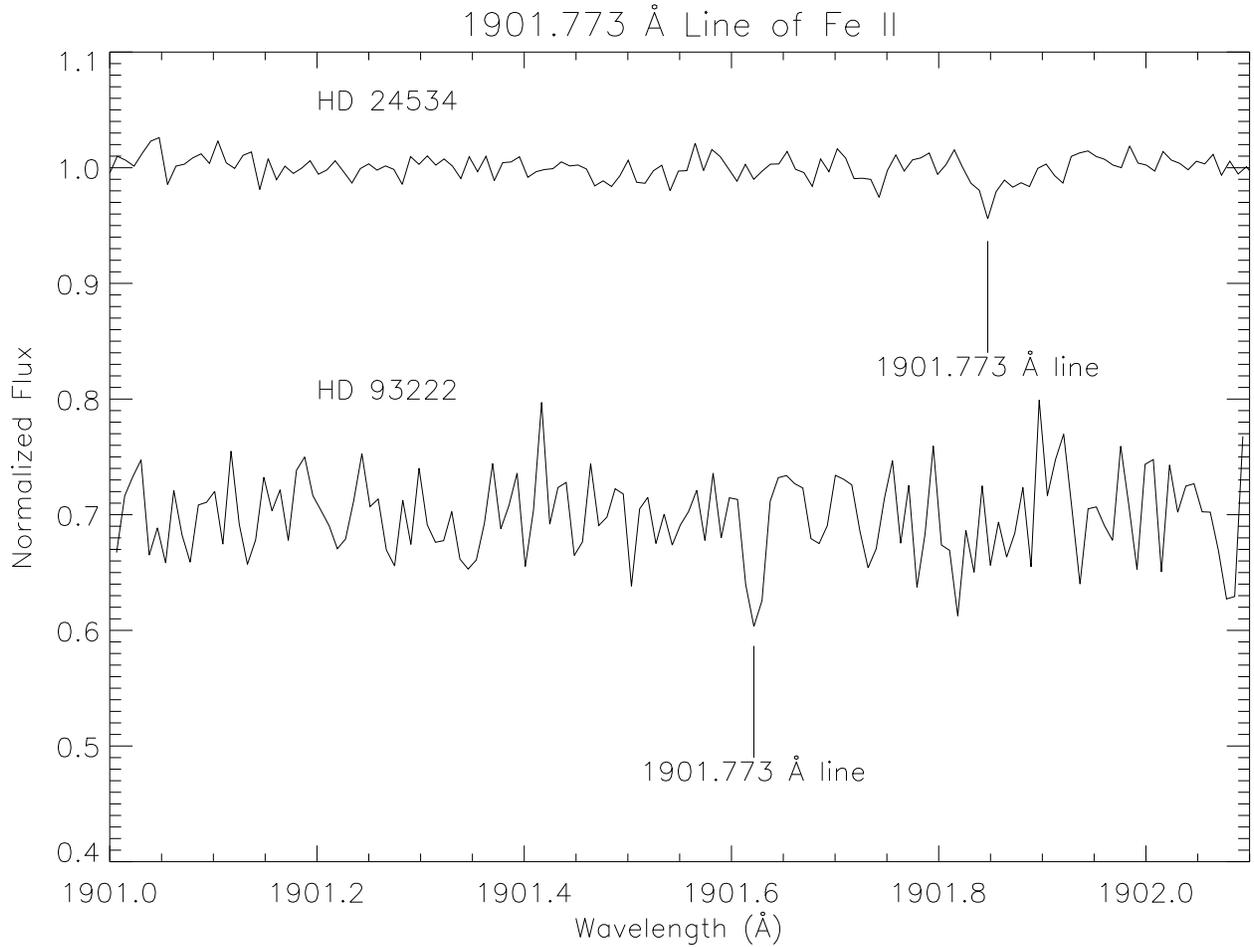}
\end{center}
\caption[Detections of the 1901.773 \AA{} line of Fe II]{Plots of the two detections of the 1901.773 \AA{} line of Fe II, for HD 24534 and HD 93222.  The HD 93222 spectrum is shifted downward by 30\% of the normalized continuum.  See \S \ref{ss:1902line} for a discussion of our identification of these features.}
\label{fig:1902line}
\end{figure}

\clearpage \clearpage

\begin{figure}
\begin{center}
\epsscale{1.00}
\plotone{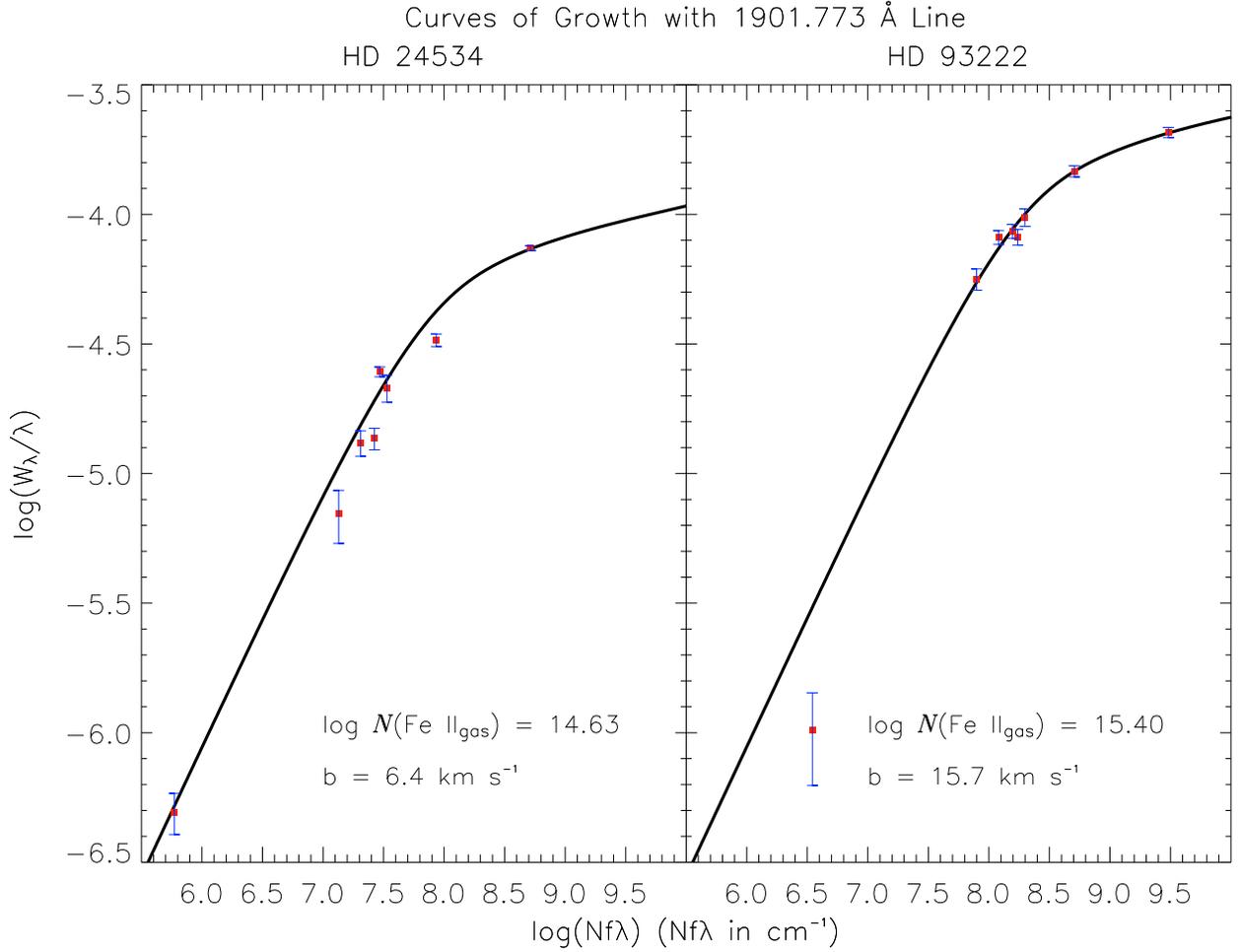}
\end{center}
\caption[Curves of Growth with a 1901.773 \AA{} Line Measurement]{Curves of growth for HD 24534 and HD 93222 when our detections of the weak 1901.773 \AA{} line are included (cf. Figures \ref{fig:cogs1-15} and \ref{fig:cogs16-30}).  The line does not strongly affect the derived column densities and $b$-values (the revised values are identical for HD 24534 and nearly identical for HD 93222).  In the case of HD 93222, there are two marginally resolved velocity components in the other lines that are seen in {\it FUSE} spectra, and the detected 1901.773 \AA{} line likely only represents one velocity component, explaining why its $\eqw$ is less than expected from the best fit for the column density.  See \S \ref{ss:1902line} for more discussion.}
\label{fig:cogs1902}
\end{figure}

\clearpage \clearpage

\begin{figure}
\begin{center}
\epsscale{1.00}
\plotone{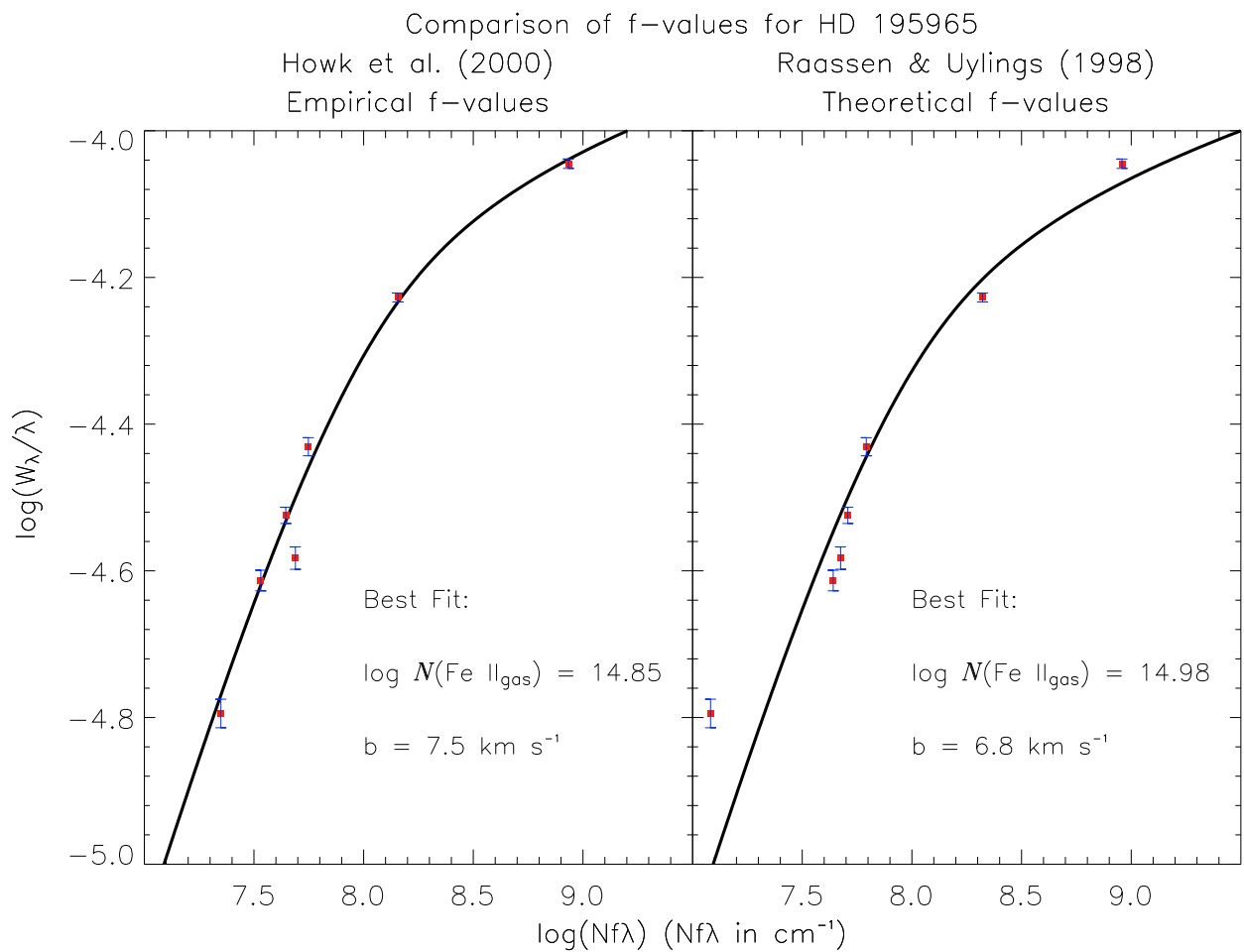}
\end{center}
\caption[Comparison of Fe II $f$-values]{A comparison of the curve-of-growth method using the empirical $f$-values of \citet{Howk} as opposed to the theoretical $f$-values of \citet{RU1998}, as applied to deriving the column density for HD 195965.  While a comprehensive statistical examination of these $f$-values is outside of the scope of this paper, we present this as an example of the extensive anecdotal evidence in this study that favors the \citet{Howk} $f$-values.  Note, however, that one discrepant line in the \citet{Howk} $f$-values is the 1112 \AA{} line.}
\label{fig:cogsfval}
\end{figure}

\clearpage \clearpage

\end{document}